\documentclass[ejs]{imsart}

\RequirePackage{amsthm,amsmath,amsfonts,amssymb}
\RequirePackage[numbers]{natbib}
\RequirePackage[colorlinks,citecolor=blue,urlcolor=blue]{hyperref}
\RequirePackage{graphicx}
\RequirePackage{bbm}
\usepackage{placeins}

\startlocaldefs

\newtheorem{theorem}{Theorem}[section]

\theoremstyle{remark}

\newtheorem{remark}{Remark}

\endlocaldefs

\begin{document}

\begin{frontmatter}
\title{Spectral graph clustering via the Expectation-Solution algorithm}
\runtitle{Spectral graph clustering via ES}
\runauthor{Z. M. Pisano et al.}

\begin{aug}
\author{\fnms{Zachary M.} \snm{Pisano}\thanksref{t1}\ead[label=e1]{zpisano1@jhu.edu}}\\
\author{\fnms{Joshua S.} \snm{Agterberg}\ead[label=e2]{jagterb1@jhu.edu}}\\
\author{\fnms{Carey E.} \snm{Priebe}\ead[label=e3]{cep@jhu.edu}}\\
\and
\author{\fnms{Daniel Q.} \snm{Naiman}\ead[label=e4]{daniel.naiman@jhu.edu}}
\address{Department of Applied Mathematics and Statistics,
Johns Hopkins University,\\
\printead{e1,e2,e3,e4}}

\thankstext{t1}{Corresponding author. This work is partially supported by the D3M program of the Defense Advanced Research Projects Agency (DARPA). The authors would like to acknowledge and thank the referees, whose commentary and criticism of an earlier version of the present article improved it considerably.}
\end{aug}

\begin{abstract}
The stochastic blockmodel (SBM) models the connectivity within and between disjoint subsets of nodes in networks. Prior work demonstrated that the rows of an SBM's adjacency spectral embedding (ASE) and Laplacian spectral embedding (LSE) both converge in law to Gaussian mixtures where the components are curved exponential families. Maximum likelihood estimation via the Expectation-Maximization (EM) algorithm for a full Gaussian mixture model (GMM) can then perform the task of clustering graph nodes, albeit without appealing to the components' curvature. Noting that EM is a special case of the Expectation-Solution (ES) algorithm, we propose two ES algorithms that allow us to take full advantage of these curved structures. After presenting the ES algorithm for the general curved-Gaussian mixture, we develop those corresponding to the ASE and LSE limiting distributions. Simulating from artificial SBMs and a brain connectome SBM reveals that clustering graph nodes via our ES algorithms can improve upon that of EM for a full GMM for a wide range of settings.
\end{abstract}



\end{frontmatter}

\section{Introduction}

Statistical inference on graphs is a burgeoning field of study in statistics with applications in neuroscience \citep{Lys-2017, Priebe-2019} and social networks \cite{Karrer-2011}, among other areas of research. Given a random graph $G$ on $n$ vertices generated by some model $F$, a typical procedure is to embed its adjacency matrix $\textbf{A} \in \{0,1\}^{n \times n}$ into a lower dimensional space $\mathbbm R^d$ as a collection of $n$ points. Commonly chosen embeddings include the \textit{adjacency spectral embedding} (ASE) and \textit{Laplacian spectral embedding} (LSE), obtained via the truncated eigendecomposition of $\textbf{A}$ and its normalized Laplacian $\boldsymbol{\mathcal L}\textbf{(A)}$. From here one may seek to perform the task of clustering these points and --- by extension --- their corresponding vertices.

The spectral graph clustering problem has been extensively studied for settings in which the graph (or graphs) on hand are posited to have been generated by a stochastic blockmodel (SBM) \cite{SBM-83}, with many results regarding consistent recovery of the block assignments being known \cite{Fishkind-13}. \cite{Avanti-16} and \cite{Tang-18} showed that the distribution of the points of the ASE and LSE both converge to curved normal mixture distributions. Moreover, they demonstrated that clustering via the Expectation-Maximization (EM) algorithm \cite{DLR} for a Gaussian mixture model (GMM) performs better than doing so via the $K$-means algorithm. However, their implementation of the EM algorithm failed to take into account the curved structure of the mixture's component distributions, and may have therefore resulted in an increased number of clustering errors.

This paper seeks to improve upon their results by introducing an Expectation-Solution (ES) algorithm \cite{ES-algorithm} that makes full use of the ASE and LSE limiting distributions' curved-normal structure. We accomplish this by noting that the complete-data likelihood equations from the EM algorithm may be reinterpreted as complete-data estimating equations used by ES. Since each component of a curved-Gaussian mixture model (C-GMM) encodes its variance as a function of the component means and potentially the mixing proportions, the estimating equations corresponding to the component variances are rendered superfluous. This leads to an iterative scheme in which the component means and mixing proportions are updated as in EM for a full GMM, but the component variances are updated by simply plugging in the new means and proportions into the variance functions.

To the best of our knowledge this paper represents the first time the ES algorithm has been utilized for a C-GMM, and owing to its relative simplicity we cannot help but recommend it as a novel tool for the practitioner's toolbox, one that is not only limited in its usefulness to the spectral clustering problem. We structure the rest of the paper as follows: In Section \ref{background_setting} we review the setting and background for random dot product graphs and SBMs. In Section \ref{ES_algorithms} we present the ES algorithm, first for a generic C-GMM, and then for both the ASE and LSE limiting distributions. In Sections \ref{sim} and \ref{discussion} we present and discuss the results of our algorithm for simulated data from artificial SBMs, a brain connectome SBM, and a rank-deficient SBM.

\subsection{Notation}
Except where otherwise specified we use emboldened capital letters such as $\textbf{A},\textbf{B},\textbf{X}$, etc. to denote matrices, emboldened and italicized capital letters with one indexed subscript to denote a row of the corresponding matrix as a column vector, as in $\boldsymbol A_i$, $\boldsymbol B_j$, $\boldsymbol X_k$. Emboldened lowercase Greek letters $\boldsymbol{\pi}, \boldsymbol{\tau}$ are vectors. We take $\boldsymbol\Delta^K := \{(\pi_1,\pi_2, \dots, \pi_K)\in \mathbbm R_{\geq 0}^K \ | \ \sum_{k=1}^K\pi_k =1\}$ to be the unit simplex in $\mathbbm R^K$, and $\delta_c$ to be the probability distribution assigning point mass to its parameter $c$ in Euclidean space. The vector of all ones and all zeros in $\mathbbm R^n$ are given by $\boldsymbol 1_n$ and $\boldsymbol 0_n$, respectively. We omit subscripts where the dimension is obvious.

Given a real symmetric $n \times n$ matrix $\textbf{M}$ we write its spectral decomposition as \begin{equation*}\textbf{M} = \textbf{U}_{\textbf{M}}\textbf{D}_{\textbf{M}}\textbf{U}_{\textbf{M}}^\top,\end{equation*} where $\textbf{U}_{\textbf{M}}$ is a unitary matrix and $\textbf{D}_{\textbf{M}} = \text{diag}(\lambda_1^{(\textbf{M})}, \lambda_2^{(\textbf{M})}, \dots, \lambda_n^{(\textbf{M})})$ such that $\lambda_1^{(\textbf{M})} \geq \lambda_2^{(\textbf{M})} \geq \cdots \geq \lambda_n^{(\textbf{M})}$ are the ordered eigenvalues of $\textbf{M}$. We define the normalized Laplacian of $\textbf{M}$ as $\boldsymbol{\mathcal L}\textbf{(M)} = (\text{diag}(\textbf{M}\boldsymbol 1_n)^{-\frac{1}2}) \textbf{M}(\text{diag}(\textbf{M}\boldsymbol 1_n)^{-\frac{1}2})$, with spectral decomposition $\textbf{\~ U}_{\textbf{M}} \textbf{\~ D}_{\textbf{M}}\textbf{\~ U}_{\textbf{M}}^\top$.

In Section \ref{sim} we use the shorthand \begin{equation}\label{methods}\{K\text{-means}, EM, ES\} \circ \{ASE, LSE\}\end{equation} to refer to each clustering method; e.g., $EM\circ ASE$ is to be read as ``EM for the ASE." Finally, in sections where we are explicitly concerned with GMMs, any reference to the EM algorithm is understood to refer to the EM algorithm for a full GMM as outlined in \cite{Raftery-02}.

\section{Background and Setting}\label{background_setting}

The $K$-block SBM encodes the probabilistic connectivity within and between disjoint subsets of graph nodes. For $K \geq 2$, $\boldsymbol\pi \in \boldsymbol \Delta^K$, and $\textbf{B} \in (0,1)^{K\times K}$ a rank-$d$ symmetric matrix with distinct rows, we write $(\textbf{A}, \boldsymbol\tau) \sim $ SBM$(n, \textbf{B}, \boldsymbol\pi)$ with sparsity factor $\rho_n \in (0,1\rbrack$ provided the following:
\begin{align*}
\boldsymbol\tau &:= (\tau_1, . . ., \tau_n) \\
\tau_i &\overset{i.i.d.}\sim \text{Cate}(\boldsymbol\pi) \\
A _{ij} \ | \ \boldsymbol\tau &\overset{ind.}\sim \text{Bern}(\rho_n B_{\tau_i\tau_j}),\ i < j\\
A_{ij} &= A_{ji}\\
\text{diag}(\textbf{A}) &= \boldsymbol 0_n.
\end{align*}
Note $\boldsymbol\tau$ denotes the vector of block memberships. If only $\textbf{A}$ is observed, we write $\textbf{A} \sim $ SBM$(n, \textbf{B}, \boldsymbol\pi)$.

The sparsity factor $\rho_n$ indexes a sequence of models where the edge probabilities change with $n$. We mention the sparsity factor purely for the sake of completeness, but common assumptions include taking $\rho_n \equiv 1$ for all $n$ or $\rho_n\to 0$ such that $\rho_n > \log^4n/n$. The former assumption is equivalent to the assumption that there exists $c>0$ such that $\rho_n \to c$; \cite{Tang-18} used the latter assumption to establish concentration in spectral norm of $\textbf{A}$ and $\boldsymbol{\mathcal L}\textbf{(A)}$ around $\rho_n\textbf{P}$ and $\boldsymbol{\mathcal L}\textbf{(P)}$(where $\textbf{P}$ is discussed below). As we are primarily concerned with spectral clustering of the nodes of a single graph at a time, the sparsity factor is taken to be identically 1, and thus suppressed, throughout the remainder of the paper.

The SBM is a special case of the random dot product graph (RDPG) \cite{Sussman-2012}, which encodes the probability that two nodes in a random graph share an edge as defined by the $n \times n$ symmetric edge probability matrix $\textbf{P}$ such that $A_{ij} \overset{ind.}\sim \text{Bern}(P_{ij})$ for $i < j$ \cite{Young-07}. In lieu of an edge probability matrix, one may instead define a random graph model via $\textbf{X}\in \mathbbm R^{n\times d}$ such that the row magnitudes are bounded above by 1 and the dot product between any two rows falls within the unit interval. In such case we permit the mathematical convenience of allowing for self loops, leading us to formulate a random graph model with edge probability matrix $\textbf{P} = \textbf{XX}^\top$. We refer to the rows of $\textbf{X}$ as the \textit{latent positions} of the graph model. Note that the latent positions are inherently unidentifiable, since for any $d\times d$ orthogonal matrix $\textbf{W}$ we have that $(\textbf{XW})(\textbf{XW})^\top = \textbf{XX}^\top$. One arrives at an analogous means of defining a rank-$d$, $K$-block SBM by defining $\boldsymbol \nu_1, \dots, \boldsymbol\nu_K \in \mathbbm R^d$ such that $\|\boldsymbol \nu_k\|\leq 1$ for all $k$ and $\boldsymbol\nu_k^\top \boldsymbol\nu_{k^{'}} \in (0,1)$ for each $k,k^{'}$ pair. If we define $\textbf{x} \in \mathbbm R^{K\times d}$ such that the $k$th row is $\boldsymbol\nu_k^\top$, then $\textbf{xx}^\top$ is the block probability matrix of an SBM, with latent positions $\boldsymbol \nu_k.$

For $d\leq n$, the $d$-dimensional \textit{adjacency spectral embedding} (ASE) of $\textbf{A}$ is given by the $n\times d$ matrix $\textbf{\^ X} := \textbf{U}_{\textbf{A}}^{(d)}(\textbf{D}_{\textbf{A}}^{(d)})^{\frac12}$, where the columns of $\textbf{U}_{\textbf{A}}^{(d)}$ are the first $d$ columns of $\textbf{U}_{\textbf{A}}$ and $\textbf{D}_{\textbf{A}}^{(d)}$ is the $d\times d$ principal minor of $\textbf{D}_{\textbf{A}}$ consisting of the $d$ largest ordered eigenvalues on the diagonal \cite{ASE-14}. The $n$ rows of $\textbf{\^ X}$ can thus be thought of as a collection of points in $\mathbbm R^d$ that estimate the true latent positions up to orthogonal transformation. We can, therefore, touch upon the notion of how a random choice of an unobservable latent position $\boldsymbol X_i$ (say, via some distribution $F$ on $\mathbbm R^d$), informs the distribution of the corresponding estimated latent position $\boldsymbol{\hat X}_i$. For a $K$-block SBM we have that the distribution $F$ on the latent positions is a mixture of point masses, with mixture weights $\pi_k$. \cite{Avanti-16} obtained the following result as a corollary to a more general theorem regarding RDPGs and SBMs with $\rho_n \to 0$ such that $\rho_n > \log^4n/n$.

\begin{theorem}\label{thm:ase}
Let $\boldsymbol X_i \overset{i.i.d.}\sim F = \sum_{k=1}^K\pi_k\delta_{\boldsymbol\nu_k}$, where $\boldsymbol\pi \in \boldsymbol \Delta^K$. Define $\boldsymbol\Lambda = \mathbbm E \lbrack \boldsymbol X_1 \boldsymbol X_1^\top  \rbrack$. Also let $\boldsymbol \Sigma(\boldsymbol x) = \boldsymbol\Lambda^{-1}\mathbbm E \lbrack \boldsymbol X_1 \boldsymbol X^\top _1(\boldsymbol x^\top \boldsymbol X_1-\boldsymbol x^\top \boldsymbol X_1\boldsymbol X^\top _1 \boldsymbol x) \rbrack \boldsymbol\Lambda^{-1}$. If $\rho_n \equiv 1$, there exists a sequence of orthogonal matrices $\textbf W_n$ such that for any fixed index $i$
\begin{equation*}
\sqrt{n}(\textbf W_n \boldsymbol{\hat X}_i - \boldsymbol X_i) | \boldsymbol X_i = \boldsymbol\nu_k \overset{d}\to \mathcal N(\boldsymbol 0, \boldsymbol \Sigma(\boldsymbol\nu_k)).
\end{equation*}
\end{theorem}

That is to say that the optimally rotated rows $\boldsymbol{\hat X}_i$ of the ASE $\textbf{\^ X}$ for a graph generated by an SBM are approximately distributed as a mixture of curved multivariate normal distributions, each centered at the corresponding true latent positions $\boldsymbol X_i$ of $\textbf{P}$. As mentioned above, we reiterate that these true latent positions are assumed to arise from a mixture of point masses on the scaled eigenvectors of $\textbf{B}$. Owing to the method by which the ASE is computed, we note that the rows are identically distributed but not independent.

The \textit{Laplacian spectral embedding} (LSE) is to the normalized Laplacian $\boldsymbol{\mathcal L}\textbf{(A)}$ as the ASE is to $\textbf{A}$. That is, for $d \leq n$ the $d$-dimensional LSE is given by $\textbf{\v X} = \textbf{\~ U}_{\textbf{A}}^{(d)}(\textbf{\~ D}_{\textbf{A}}^{(d)})^{\frac12}$, where $\textbf{\~ U}_{\textbf{A}}^{(d)}$ and $\textbf{\~ D}_{\textbf{A}}^{(d)}$ are analogous to $\textbf{U}_{\textbf{A}}^{(d)}$ and $\textbf{D}_{\textbf{A}}^{(d)}$. Like $\textbf{\^ X}$, the $n$ rows of $\textbf{\v X}$ can be thought of as a collection of points in $\mathbbm R^d$. The LSE is preferred in settings where the adjacency matrix is sparse or the edge probability matrix is believed to be of the form $\rho_n \textbf{P}_n$ for some positive sequence $\rho_n \to 0$ as $n\to\infty$, since the normalized Laplacian is the same for $\textbf{P}$ as it is for $\rho \textbf{P}$; that is,
\begin{align*}
\boldsymbol{\mathcal L(}\rho \textbf{P)} &= (\text{diag}((\rho \textbf{P})\boldsymbol 1_n)^{-\frac{1}2}) (\rho \textbf{P})(\text{diag}((\rho \textbf{P})\boldsymbol 1_n)^{-\frac{1}2})\\ &= \rho^{-1} \rho (\text{diag}(\textbf{P}\boldsymbol 1_n)^{-\frac{1}2}) \textbf{P}(\text{diag}( \textbf{P}\boldsymbol 1_n)^{-\frac{1}2})\\
&= \boldsymbol{\mathcal L}\textbf{(P)}.
\end{align*}
\cite{Tang-18} obtained the following result for the LSE of an SBM, analogous to the previous theorem.

\begin{theorem}\label{thm:lse}
Let the setting be as in Theorem \ref{thm:ase}, and also let $\boldsymbol\mu = \mathbbm E\lbrack \boldsymbol X_1\rbrack$ and
\begin{equation*}
\boldsymbol{\tilde\Lambda} = \mathbbm E \bigg\lbrack \frac{\boldsymbol X_1 \boldsymbol X_1^\top }{\boldsymbol X_1^\top \boldsymbol\mu}\bigg\rbrack.
\end{equation*} 
Define 
\begin{equation*}
\boldsymbol{\tilde\Sigma}(\boldsymbol x) = \mathbbm E \bigg\lbrack \bigg(\frac{\boldsymbol{\tilde\Lambda}^{-1}\boldsymbol X_1}{\boldsymbol X_1^\top \boldsymbol\mu} - \frac{\boldsymbol x}{2\boldsymbol x^\top \boldsymbol\mu} \bigg ) \bigg ( \frac{\boldsymbol X_1^\top \boldsymbol{\tilde\Lambda}^{-1}}{\boldsymbol X_1^\top \boldsymbol\mu} - \frac{\boldsymbol x^\top }{\boldsymbol x^\top \boldsymbol\mu} \bigg)\bigg(\frac{\boldsymbol x^\top \boldsymbol X_1-\boldsymbol x^\top \boldsymbol X_1 \boldsymbol X_1^\top \boldsymbol x}{\boldsymbol x^\top \boldsymbol\mu} \bigg)\bigg\rbrack.
\end{equation*}
If $\rho_n \equiv 1$ and $n_k = | \lbrace i\leq n | X_i = \boldsymbol\nu_k\rbrace |$, then there exists a sequence of orthogonal matrices $\textbf W_n$ such that for any fixed index $i$
\begin{equation*}
n\bigg(\textbf W_n \check X_i - \frac{\boldsymbol\nu_k}{\sqrt{\sum_l n_l \boldsymbol\nu_k^\top \boldsymbol\nu_l}}\bigg) | X_i = \boldsymbol\nu_k \overset{d}\to \mathcal N(0, \boldsymbol{\tilde\Sigma}(\boldsymbol\nu_k)).
\end{equation*}
\end{theorem}

The limiting distributions justify the fitting of full GMMs to cluster the rows of either spectral embedding. Doing so improves upon the clustering performance of $K$-means. Although $K$-means is invariant to orthogonal transformations of the data, the procedure imposes the assumption of equal, spherical covariances on the clusters; however, the ASE and LSE clusters can be non-spherical and of varying spread as demonstrated in Figures \ref{fig:low_rank_ase_plot}--\ref{fig:low_rank_lse_plot} below, as well as in \cite{GRDPG} and \cite{Priebe-2019}. By contrast, clustering via the EM algorithm accounts for the possibility of unequal covariances; however, the curvature of the components remains unaddressed.

In practice, both rank$(\textbf{B})$ and $K$ are unknown. A principled method of estimating the former is to inspect the scree-plot of the singular values of $\textbf{A}$ and look for ``elbows'' defining the cut-off between singular values corresponding to signal dimensions and those corresponding to noise dimensions \cite{Droso-2017}. One can estimate the latter by means of maximizing a fitness criterion penalized by model complexity, \`a la the Akaike Information Criterion (AIC) \cite{AIC} and the Bayesian Information Criterion (BIC) \cite{BIC}; however, recent work due to \cite{yang-21} has called into question the validity and applicability of these classical criteria to SBM settings.

For a comprehensive look at the proofs of Theorems \ref{thm:ase} and \ref{thm:lse}, as well as fuller details of the setting we describe, we refer the reader to \cite{JMLR}.

\section{The Expectation-Solution Algorithms}\label{ES_algorithms}

First introduced by \cite{ES-algorithm}, the Expectation-Solution (ES) algorithm arises as a generalization of the EM algorithm \cite{DLR}, where instead of updating the parameter estimates according to a collection of complete-data likelihood equations, we update the parameters as solutions to complete-data estimating equations. When the complete-data estimating equations for a given model coincide exactly with the complete-data likelihood equations for the same model, then the ES and EM algorithms are equivalent. Thus, any comparisons drawn between EM and ES for a particular setting are actually between two different ES algorithms, a fact which allows us to use comparable convergence criteria. We will demonstrate that implementation of ES in a curved-Gaussian mixture setting allows us to make full use of the mixture components' curved structure without sacrificing the relative simplicity of EM. We review ES for the general incomplete-data setting in Appendix \ref{app:appendix1}.

\subsection{The ES Algorithm for Mixtures of Curved Gaussians}
Let us ignore, for the moment, the content of Section \ref{background_setting} and consider a generic $K$-component Gaussian mixture model (GMM) supported on $\mathbbm R^d$: \begin{equation*}\boldsymbol X_i\overset{i.i.d.}\sim\\\sum_{k=1}^K\pi_k\mathcal N( \boldsymbol \mu_k, \boldsymbol\Sigma_k), i = 1, . . .,n;\end{equation*} such that $\boldsymbol\pi:=(\pi_1, . . ., \pi_K)\in\boldsymbol \Delta^K$ and $(\boldsymbol \mu_k, \boldsymbol\Sigma_k)\in \mathbbm R^d \times \mathbbm M_d$ for all $k$, where $\mathbbm M_d$ is the space of $d\times d$ positive definite matrices. Maximum likelihood estimation of the parameters upon observing $\boldsymbol x_1, . . ., \boldsymbol x_n$ can be handily accomplished by the EM algorithm. Here, the component distributions are by themselves smooth exponential family distributions, a fact that allows us to utilize various results related to the computation of maximum likelihood estimates in the M-step.

Suppose instead that our model is
\begin{align*}
&\boldsymbol X_i \overset{i.i.d.}{\sim}\sum_{k=1}^K\pi_k\mathcal N(\boldsymbol \mu_k, \boldsymbol\Sigma_k(\boldsymbol \mu_k))\\
&\boldsymbol \mu_k \in \mathcal M \subset \mathbbm R^d \\
&\boldsymbol\Sigma_k : \mathcal M \to \mathbbm M_d, \forall k
\end{align*}
and each $\boldsymbol\Sigma_k$ is continuous and differentiable. Here each component distribution is a curved exponential family distribution, where the component variances are functions of the means \cite{Bic-Dok}; hence we refer to such a setting as a curved-Gaussian mixture model (C-GMM). If we attempt to derive complete-data likelihood equations necessary to implement an EM algorithm, we may become swiftly enmired in non-linear equations which we must solve to update the means at each iteration, depending on the structure of each $\boldsymbol\Sigma_k$.

Here the ES algorithm circumvents the potential difficulty posed by the variance functions. We begin by considering the natural unobserved data extension of $\textbf X = (\boldsymbol X_1, . . ., \boldsymbol X_n)$, which should come as no surprise to those familiar with the EM algorithm:
\begin{align*}
	\boldsymbol Z_i = (Z_{i1}, . . .,Z_{iK}) \overset{i.i.d.}\sim \text{Multinomial}(1, K, \boldsymbol\pi)\\
	\boldsymbol X_i \ |\ Z_{ik} = 1 \sim \mathcal N(\boldsymbol \mu_k, \boldsymbol\Sigma_k(\boldsymbol \mu_k)).
\end{align*}
Let $\boldsymbol\Psi$ denote the parameter vector consisting of $\pi_1,\dots,\pi_{K-1}$ and the entries of $\boldsymbol \mu_1,\dots,\boldsymbol \mu_K$. In the complete-data setting $(\textbf X, \textbf Z)$ we immediately note the following:
\begin{align*}
	\mathbbm E_{\boldsymbol\Psi} \lbrack Z_{ik} \rbrack &= \pi_k \\
	\mathbbm E_{\boldsymbol\Psi} \lbrack Z_{ik}\boldsymbol X_i \rbrack &= \pi_k\boldsymbol \mu_k,
\end{align*}
which give rise to the natural estimating equations
\begin{align*}
	\frac{\sum_{i=1}^n Z_{ik}}{n} - \pi_k &= 0\\
	\frac{\sum_{i=1}^n Z_{ik}\boldsymbol X_i}{n} - \pi_k\boldsymbol \mu_k &= 0
\end{align*}
solved by
\begin{align*}
	\hat\pi_k &= \frac{\sum_{i=1}^n Z_{ik}}{n}; \\
	\boldsymbol{\hat\mu_k} &=\frac{\sum_{i=1}^n Z_{ik}\boldsymbol X_i}{\sum_{i=1}^n Z_{ik}}.
\end{align*}
The complete-data estimating equation $U_c((\textbf X, \textbf Z), \boldsymbol\Psi)= \boldsymbol C_{\boldsymbol{\Psi}} S(\textbf X, \textbf Z) + b_{\boldsymbol{\Psi}}(\textbf X)=0$ (Appendix \ref{app:appendix1}), then, is characterized by
\begin{align*}
	\boldsymbol C_{\boldsymbol\Psi} &= \textbf{I}_{(d+1)K-1} \\
S(\textbf X, \textbf Z) &= \frac{1}{n}\begin{bmatrix}
		\sum_{i=1}^n Z_{i1} \\
		\vdots \\
		\sum_{i=1}^n Z_{i(K-1)} \\
		\sum_{i=1}^n Z_{i1}\boldsymbol X_i \\
		\vdots \\
		\sum_{i=1}^n Z_{iK}\boldsymbol X_i
\end{bmatrix}\\
	b_{\boldsymbol\Psi}(\textbf X) &= -\begin{bmatrix}
		\pi_1 \\
		\vdots \\
		\pi_{K-1} \\
		\pi_1\boldsymbol \mu_1 \\
		\vdots \\
		(1-\sum_{k=1}^{K-1}\pi_k)\boldsymbol \mu_K
	\end{bmatrix}.
\end{align*}

In computing $h(\boldsymbol\Psi | \boldsymbol{\Psi^*}) = \mathbbm E_{\boldsymbol{\Psi^*}} \lbrack S(\textbf X, \textbf Z) | \textbf X = \textbf x \rbrack$, we find that we must only obtain
\begin{align*}
	Z_{ik}^* &:= \mathbbm E_{\boldsymbol{\Psi^*}} \lbrack Z_{ik} | \boldsymbol X_i = \boldsymbol x_i \rbrack \\
	&= \frac{\pi_k^*\phi(\boldsymbol x_i | \boldsymbol \mu_k^*, \boldsymbol\Sigma_k(\boldsymbol \mu_k^*))}{\sum_{j=1}^K \pi_j^*\phi(\boldsymbol x_i | \boldsymbol \mu_j^*, \boldsymbol\Sigma_j(\boldsymbol \mu_j^*))}.
\end{align*}
Thus $h(\boldsymbol\Psi | \boldsymbol{\Psi^*}) = S(\textbf x, \textbf Z^*)$ by linearity of expectation. The next iterate $\boldsymbol{\hat\Psi}$ is then obtained by solving $\boldsymbol C_{\boldsymbol\Psi} S(\textbf x, \textbf Z^*) + b_{\boldsymbol\Psi}(\textbf X) = 0$, which yields
\begin{align*}
	\hat\pi_k &= \frac{\sum_{i=1}^n Z_{ik}^*}{n}; \\
	\boldsymbol{\hat\mu_k} &=\frac{\sum_{i=1}^n Z_{ik}^*\boldsymbol x_i}{\sum_{i=1}^n Z_{ik}^*}.
\end{align*}
We present this procedure written concisely as Algorithm \ref{alg:es_curved}.

\noindent\fbox{\parbox{\textwidth}{
\noindent\fbox{\parbox{4.72in}{\textbf{Algorithm \ref{alg:es_curved}: The ES Algorithm for the Curved-Gaussian Mixture Setting}}}\label{alg:es_curved}

1. Initialize $\boldsymbol{\Psi^*} = \boldsymbol{\Psi_0}$.

2. \textbf{E-Step:} For each $i$ and $k$ compute 
\begin{equation*}
	Z_{ik}^* = \frac{\pi_k^*\phi(\boldsymbol x_i | \boldsymbol \mu_k^*, \boldsymbol\Sigma_k(\boldsymbol \mu_k^*))}{\sum_{j=1}^K \pi_j^*\phi(\boldsymbol x_i | \boldsymbol \mu_j^*, \boldsymbol\Sigma_j(\boldsymbol \mu_j^*))}.
\end{equation*}

3. \textbf{S-Step:} Compute the entries of $\boldsymbol{\hat\Psi}$ as 
\begin{align*}
	\hat\pi_k &= \frac{\sum_{i=1}^n Z_{ik}^*}{n} \\
	\boldsymbol{\hat\mu_k} &=\frac{\sum_{i=1}^n Z_{ik}^*\boldsymbol x_i}{\sum_{i=1}^n Z_{ik}^*}.
\end{align*}

4. Take $\boldsymbol{\Psi^*} = \boldsymbol{\hat\Psi}$.

5. Repeat steps 2--4 until some convergence criterion is satisfied.
}
}
\\

Note that the only difference between this ES algorithm and the usual EM algorithm for GMMs is that we ``update" the component variances by plugging the usual updates for the component means into the respective $\boldsymbol \Sigma_k(\cdot)$ instead of computing
\begin{equation*}
	\boldsymbol{\hat\Sigma}_k = \frac{\sum_{i=1}^n Z^*_{ik}(\boldsymbol x_i - \boldsymbol{\hat\mu_k})(\boldsymbol x_i -\boldsymbol{\hat\mu_k})^\top }{\sum_{i=1}^n Z^*_{ik}}
\end{equation*}
for each $k$. As a result we estimate only $(d+1)K-1$ parameters, whereas in the full GMM setting we estimate another $K (d+ {d\choose 2})$ parameters comprising the component covariances. Since the model complexity penalties used to compute both AIC and BIC increase in magnitude with the number of parameters, the severe reduction in the number of parameters needed to be estimated in a C-GMM can vastly decrease these penalties. That is, if we are deciding between a GMM and C-GMM with approximately equal likelihoods for a given collection of data, then it is clear that the C-GMM will achieve the higher (and therefore more desirable) AIC or BIC. 

If we are interested in clustering the observations we perform the usual GMM clustering procedure, which is to assign each observation to the cluster with the highest posterior probability.

\begin{remark}
	We note here the distinction between two classes of C-GMMs. We call a C-GMM \textbf{separable} if each component variance is solely a function of its respective mean and possibly its mixing proportion. Likewise, if one or more of the variances take other components' means or mixing proportions as arguments, then that C-GMM is \textbf{tied} --- as in the cases of both spectral embeddings' limiting distributions. 
\end{remark}

\subsection{ES Algorithm for the ASE}
Let us return to the content of Section \ref{background_setting}, and assume the setting of Theorem \ref{thm:ase} with $\textbf{B} = \textbf{xx}^\top $, where the $k^\text{th}$ row of $\textbf x$ is $\boldsymbol\nu_k$, and that we have observed $\textbf{A}$ and computed its $d$-dimensional ASE $\textbf{\^ X}$. Since $F = \sum_{k=1}^K\pi_k\delta_{\boldsymbol\nu_k}$, we have that $\boldsymbol\Lambda = \sum_{k=1}^K \pi_k \boldsymbol\nu_k \boldsymbol\nu_k^\top $ and 
\begin{equation*}
	\boldsymbol\Sigma(\boldsymbol\nu_k) = \boldsymbol\Lambda^{-1}\bigg (\sum_{j=1}^K \pi_j \boldsymbol\nu_j \boldsymbol\nu_j^\top (\boldsymbol\nu_k^\top \boldsymbol\nu_j-\boldsymbol\nu_k^\top \boldsymbol\nu_j \boldsymbol\nu_j^\top \boldsymbol\nu_k)\bigg)\boldsymbol\Lambda^{-1}.
\end{equation*}
The variance function thus takes every latent position and mixture weight as arguments, and our iterative scheme must reflect this. Therefore, let $\boldsymbol\Sigma(\cdot | \textbf x, \boldsymbol\pi)$ denote the covariance function instead. The normal mixture model we thus consider is
\begin{equation*}
\boldsymbol{\hat X}_i \sim \sum_{k=1}^K \pi_k \mathcal N\bigg(\boldsymbol\nu_k,\frac{\boldsymbol\Sigma(\boldsymbol\nu_k|\textbf x, \boldsymbol\pi)}{n}\bigg),
\end{equation*}
where each $\boldsymbol{\hat X}_i$ is identically distributed. Note that we have dropped the optimal orthogonal transformation $\textbf{W}_n$ in the statement of Theorem 2.1; hence the algorithm (presented below as Algorithm \ref{alg:es_ase}) estimates the latent positions up to this rotation, but this has no bearing on the task of clustering the graph nodes or estimating the block probability matrix and membership probabilities.

\noindent\fbox{\parbox{\textwidth}{
\noindent\fbox{\textbf{Algorithm \ref{alg:es_ase}: The ES Algorithm for the ASE}}\label{alg:es_ase}

1. Initialize $\boldsymbol{\Psi^*} = \boldsymbol{\Psi_0}$.

2. \textbf{E-Step:} Compute
\begin{equation*}
Z^*_{ik} =\frac{\pi^*_k \phi(\boldsymbol{\hat X}_i | \boldsymbol\nu^*_k, \boldsymbol\Sigma(\boldsymbol\nu^*_k|\textbf x^*, \boldsymbol \pi^*)/n)}{\sum_{j=1}^K\pi^*_j \phi(\boldsymbol{\hat X}_i | \boldsymbol\nu_j^*, \boldsymbol\Sigma(\boldsymbol\nu_j^*|\textbf x^*, \boldsymbol \pi^*)/n)}
\end{equation*}

3. \textbf{S-Step:} Compute
\begin{align*}
\hat \pi_k &= \frac{\sum_{i=1}^nZ_{ik}^*}{n} \\
\boldsymbol{\hat\nu}_k &= \frac{\sum_{i=1}^n Z_{ik}^*\boldsymbol{\hat X}_i}{\sum_{i=1}^nZ_{ik}^*}.
\end{align*}

4. Take $\boldsymbol{\Psi^*} = \boldsymbol{\hat\Psi}$.

5. Repeat steps 2-4 until some convergence criterion is satisfied.
}
}
\begin{remark}
We can construct a separable analogue of Algorithm \ref{alg:es_ase} by updating each variance with the newest iterate of the corresponding mean and proportion while holding all other arguments as their previous iterates, only updating them every $\iota$ iterations, or holding them constant. Such schemes (particularly the last) greatly alter the model at hand, hence we would not recommend them for use; rather, we only mention them to fill in the middle ground between Algorithms \ref{alg:es_curved} and \ref{alg:es_ase}.
\end{remark}

\subsection{ES Algorithm for the LSE}
Let us assume the setting of Theorem \ref{thm:lse}. We have
\begin{align*}
\boldsymbol \mu &= \sum_{k=1}^K \pi_k\boldsymbol\nu_k \\
\boldsymbol{\tilde\Lambda} &= \sum_{k = 1}^K \pi_k\frac{\boldsymbol\nu_k\boldsymbol\nu_k^\top }{\boldsymbol\nu_k^\top \boldsymbol \mu} \\
\boldsymbol{\tilde\Sigma}(\boldsymbol\nu_k | \textbf x, \boldsymbol\pi) &= \sum_{j = 1}^K\pi_j  \bigg(\frac{\boldsymbol{\tilde\Lambda}^{-1}\boldsymbol\nu_j}{\boldsymbol\nu_j^\top \boldsymbol \mu} - \frac{\boldsymbol\nu_k}{2\boldsymbol\nu_k^\top \boldsymbol \mu} \bigg ) \bigg ( \frac{\boldsymbol\nu_j^\top \boldsymbol{\tilde\Lambda}^{-1}}{\boldsymbol\nu_j^\top \boldsymbol \mu} - \frac{\boldsymbol\nu_k^\top }{\boldsymbol\nu_k^\top \boldsymbol \mu} \bigg)\bigg(\frac{\boldsymbol\nu_k^\top \boldsymbol\nu_j-\boldsymbol\nu_k^\top \boldsymbol\nu_j\boldsymbol\nu_j^\top \boldsymbol\nu_k}{\boldsymbol\nu_k^\top \boldsymbol \mu} \bigg).
\end{align*}
The normal mixture model we consider here is 
\begin{equation*}
\boldsymbol{\check X}_i \sim \sum_{k=1}^K \pi_k\mathcal N\bigg( \frac{\boldsymbol\nu_k}{\sqrt{\sum_l n_l \boldsymbol\nu_l^\top \boldsymbol\nu_k}}, \frac{\boldsymbol{\tilde\Sigma}(\boldsymbol\nu_k| \textbf x, \boldsymbol\pi)}{n^2} \bigg ),
\end{equation*}
where each $\boldsymbol{\check X}_i$ is identically distributed.
Owing to the presence of the unobserved $n_k$ in the component means, the fact that the ASE and LSE are defined by the 1-1 transformation \begin{equation}\label{1-1}\textbf{\v X} = (\text{diag}(\textbf{A}\boldsymbol 1_n)^{-1/2})\textbf{\^ X},\end{equation} and the fact that the LSE covariance function takes the ASE component means as arguments, we cannot simply invoke an analogue of the above algorithm. We propose expanding the parameter vector $\boldsymbol\Psi$ by treating the $n_k$ as parameters that can be obtained by the $K$ estimating equations
\begin{equation*}
\mathbbm E_{\boldsymbol\Psi}\big\lbrack \sum_{i=1}^n Z_{ik} \big\rbrack - n_k = 0,
\end{equation*}
which would be solved by $\hat n_k := \sum_{i=1}^nZ_{ik} = n\hat\pi_k$ if we observed the true cluster labels of each row of $\textbf{\v X}$.

Upon observing $\textbf{A}$ and computing both $\textbf{\^ X}$ and $\textbf{\v X}$, we propose the following algorithm:

\noindent\fbox{\parbox{\textwidth}{
\noindent{\fbox{\textbf{Algorithm \ref{alg:es_lse}: The ES Algorithm for the LSE}}}\label{alg:es_lse}

1. Initialize $\boldsymbol{\Psi^*} = \boldsymbol{\Psi_0}$ and take $\boldsymbol\Sigma_k(\cdot) = \boldsymbol{\tilde\Sigma}(\cdot | \textbf x, \boldsymbol\pi)/n^2$.

2. \textbf{E-Step:} Compute
\begin{align*}
\boldsymbol \mu^*_k &=  \frac{\boldsymbol\nu^*_k}{\sqrt{\sum_l n_l^* \boldsymbol\nu_l^{*T}\boldsymbol\nu^*_k}}\\
Z^*_{ik} &=\frac{\pi^*_k \phi(\boldsymbol{\check X}_i | \boldsymbol \mu^*_k, \boldsymbol{\tilde\Sigma}(\boldsymbol\nu^*_k|\textbf x^*, \boldsymbol\pi^*)/n^2)}{\sum_{j=1}^K\pi^*_j \phi(\boldsymbol{\check X}_i | \boldsymbol \mu_j^*, \boldsymbol{\tilde\Sigma}(\boldsymbol\nu_j^*|\textbf x^*, \boldsymbol \pi^*)/n^2)}.
\end{align*}

3. \textbf{S-Step:} Compute
\begin{align*}
\hat \pi_k &= \frac{\sum_{i=1}^nZ_{ik}^*}{n} \\
\boldsymbol{\hat\nu}_k &= \frac{\sum_{i=1}^n Z_{ik}^*\boldsymbol{\hat X}_i}{\sum_{i=1}^nZ_{ik}^*}\\
\hat n_k &= n\hat\pi_k.
\end{align*}

4. Take $\boldsymbol{\Psi^*} = \boldsymbol{\hat\Psi}$.

5. Repeat steps 2-4 until some convergence criterion is satisfied.
}
}

Note that this algorithm makes full use of both the LSE and ASE; the $\boldsymbol{\check X}_i$ are used in the E-Step, but the $\boldsymbol{\hat X}_i$ are used in the S-Step. Even so, clustering is to be done based on the rows of $\textbf{\v X}$, as they correspond exactly to the posterior probabilities computed while the algorithm runs its course. This is due to the fact that in our implementation we specify that the conditional distribution of the $Z_{ik}$ depend purely on $\boldsymbol{\check X}_i$, meanwhile the estimating equations for the $\boldsymbol \nu_k$ remain as in the previous algorithm.

\begin{remark}\label{rem:complexity}
	 A major advantage of Algorithms \ref{alg:es_ase} and \ref{alg:es_ase} over the usual EM algorithm for a GMM is lower computational complexity to update the component variances at each iteration. Simple inspection of the M-step in the classical method reveals that the number of operations needed to update each of the $K$ component variances is $O(nd)$. In our ES algorithms the number of operations to perform the same task is $O(d^3K)$ for the ASE and $O(d^4K)$ for the LSE; the latter complexity may be shaved down to $O(d^3 K)$ by preserving initial moment estimates of $\boldsymbol\mu$ and $\boldsymbol{\tilde\Lambda}$ (and computing the latter's inverse) as described in Section \ref{discussion}. Since $d \ll n$ these ES algorithms possess far less complexity than the usual EM.
\end{remark}

\begin{remark}
	While Algorithms 2 and 3 do address the curvature of the spectral embeddings' limiting mixture distributions, the dependence of the estimated latent positions remains unaddressed, a shortcoming shared by the EM algorithm in these settings. Both ES and EM require the assumption of independence to carry out the E-step, since the conditional expectation of each $Z_{ik}$ is actually given all the observed data, not just the corresponding $\boldsymbol{\hat X}_i$ or $\boldsymbol{\check X}_i$.
\end{remark}

\section{Simulations}\label{sim}
For each simulation setting we generated 100 graphs and compared the performance of our ES algorithms with their respective EM analogues. The specific EM algorithm we used was the function $em$ from the R package Mclust \cite{Raftery-16}. All four procedures were initialized at the true parameter values. Since EM is a maximum likelihood procedure, the default convergence criterion is to terminate when updates to the log-likelihood are less than $1\times 10^{-5}$. By contrast, there is no convenient objective function associated with the ES algorithm, since the iterates arise as solutions to complete-data estimating equations; moreover, it was observed anecdotally in a few of our settings that the log-likelihood may decrease after an ES iteration. With that in mind, we noted that the EM algorithm actually is an ES algorithm where the complete-data estimating equations are the complete-data likelihood equations; and we altered the code of $em$ to reflect this. Therefore convergence of EM and ES for the ASE was assumed when the Euclidean distance between successive estimates of the parameter vector $\boldsymbol\Psi$ consisting of the mixing proportions and the entries of \textbf{x} was less than $1\times 10^{-6}$, and convergence of EM and ES for the LSE was assumed when the Euclidean distance between successive estimates of the parameter vector $\boldsymbol{\tilde\Psi}$ consisting of the mixing proportions and the scaled latent positions was less than $1\times 10^{-7}$ for the LSE, or the number of iterations exceeded 10,000. The lower threshold for $ES\circ LSE$ was based on the fact that in computing the LSE we effectively scale the rows of the ASE towards the origin (see, e.g., Figures \ref{fig:low_rank_ase_plot}--\ref{fig:low_rank_ase_plot} below), thereby decreasing the magnitude of the latent position iterates and the extent to which they can change significantly between each pass of the algorithms.

To evaluate clustering performance we computed the adjusted Rand index (ARI) \cite{ARI} for the cluster assignments from each method with the true cluster labels. For each clustering method $m$ found in (\ref{methods}), we let $ARI_m$ denote the ARI produced by that method. In the non-graph setting below (Subsection \ref{non_graph}) for $\ell = A, L$ we performed Wilcoxon rank-sum tests \cite{Wilcox-rank-sum}, with hypotheses
\begin{align*}
&H_o : \text{med}(ARI_{EM\circ\ell SE}) = \text{med}(ARI_{ES\circ\ell SE}) \\
&H_a : \text{med}(ARI_{EM\circ\ell SE}) \neq \text{med}(ARI_{ES\circ\ell SE}).
\end{align*} We present these in the form of 95\% confidence intervals for the median of each $\Delta_\ell := ARI_{EM\circ\ell SE}-ARI_{ES\circ\ell SE}$.

To evaluate accuracy of the parameter estimates, we computed the squared error of each method's terminating estimate from the true parameter vector $\boldsymbol \Psi$. As will be seen in Subsection \ref{balanced_affinity} we circumvented the issue of the latent positions' non-identifiability by rotating and centering our simulated data over the ``canonical'' latent postions in the first orthant by way of a Procrustes transformation. We take $\boldsymbol \Psi_{ASE}$ to consist of the entries of $\boldsymbol\pi$ and $\textbf x$, as well as those of the covariance matrices $\frac{\boldsymbol\Sigma(\boldsymbol\nu_k|\textbf x, \boldsymbol\pi)}n$. As $EM\circ LSE$ does not outright produce estimates for the $\boldsymbol\nu_k$, we take $\boldsymbol\Psi_{LSE}$ to consist of the entries of $\boldsymbol\pi$, $\boldsymbol \mu_k := \frac{\boldsymbol\nu_k}{\sqrt{\sum_l n_l\boldsymbol\nu_l^\top \boldsymbol\nu_k}}$, and the covariance matrices $\frac{\boldsymbol{\tilde\Sigma}(\boldsymbol\nu_k | \textbf x, \boldsymbol\pi)}{n^2}$. We then performed Wilcoxon rank sum tests for the paired collection of squared errors for $\{EM, ES\}\circ ASE$, as well as $\{EM, ES\}\circ LSE$. We let $p_{A}$ denote the $p$-value corresponding to the test
\begin{align*}
&H_o : \text{med}(\|\boldsymbol{\hat\Psi}_{EM\circ ASE}-\boldsymbol\Psi_{ASE}\|) \leq \text{med}(\|\boldsymbol{\hat\Psi}_{ES\circ ASE}-\boldsymbol\Psi_{ASE}\|)\\
&H_a : \text{med}(\|\boldsymbol{\hat\Psi}_{EM\circ ASE}-\boldsymbol\Psi_{ASE}\|) > \text{med}(\|\boldsymbol{\hat\Psi}_{ES\circ ASE}-\boldsymbol\Psi_{ASE}\|),
\end{align*}
and $p_{L}$ be that corresponding to the test
\begin{align*}
&H_o : \text{med}(\|\boldsymbol{\hat\Psi}_{EM\circ LSE}-\boldsymbol\Psi_{LSE}\|) \leq \text{med}(\|\boldsymbol{\hat\Psi}_{ES\circ LSE}-\boldsymbol\Psi_{LSE}\|)\\
&H_a : \text{med}(\|\boldsymbol{\hat\Psi}_{EM\circ LSE}-\boldsymbol\Psi_{LSE}\|) > \text{med}(\|\boldsymbol{\hat\Psi}_{ES\circ LSE}-\boldsymbol\Psi_{LSE}\|).
\end{align*}

\subsection{Non-Graph Setting}\label{non_graph}
We first tested $ES\circ\{ASE, LSE\}$ on random data generated directly from the mixture distributions given in Theorems 2.1 and 2.2. We considered the following models:
\begin{align*}
&\textbf x_1 = \begin{bmatrix}0.6210 &0.3382\\0.3382 &0.6210\end{bmatrix}
&\textbf x_2 = \begin{bmatrix}0.4076 &0.1840\\0.1840 &0.4076\end{bmatrix}\\
&\textbf x_3 = \begin{bmatrix}0.6024 &0.3703\\0.3703 &0.5319\end{bmatrix}
&\textbf x_4 = \begin{bmatrix}0.3962 &0.2074\\0.2074 &0.3721\end{bmatrix}
\end{align*}
with $\boldsymbol\pi = (\frac 12,\frac 12)$ in all cases. These particular choices of $\textbf x$ correspond to, respectively, balanced affinity models given by $(a,b) = (0.5, 0.4)$ and $(a,b) = (0.2, 0.15)$, and balanced core-periphery models given by $(a,b) = (0.5, 0.42)$ and $(a,b) = (0.2, 0.15)$ which are included in the simulations in the following two subsections. Each randomly generated sample was taken to comprise $\textbf{\^ X}$ and the simulated LSE $\textbf{\v X}$ was computed via the 1--1 relationship (\ref{1-1}), taking $\textbf{A} = \textbf{\^ X} \textbf{\^ X}^\top $.

\begin{figure}[h!]
	\centering
	\includegraphics[width = \textwidth]{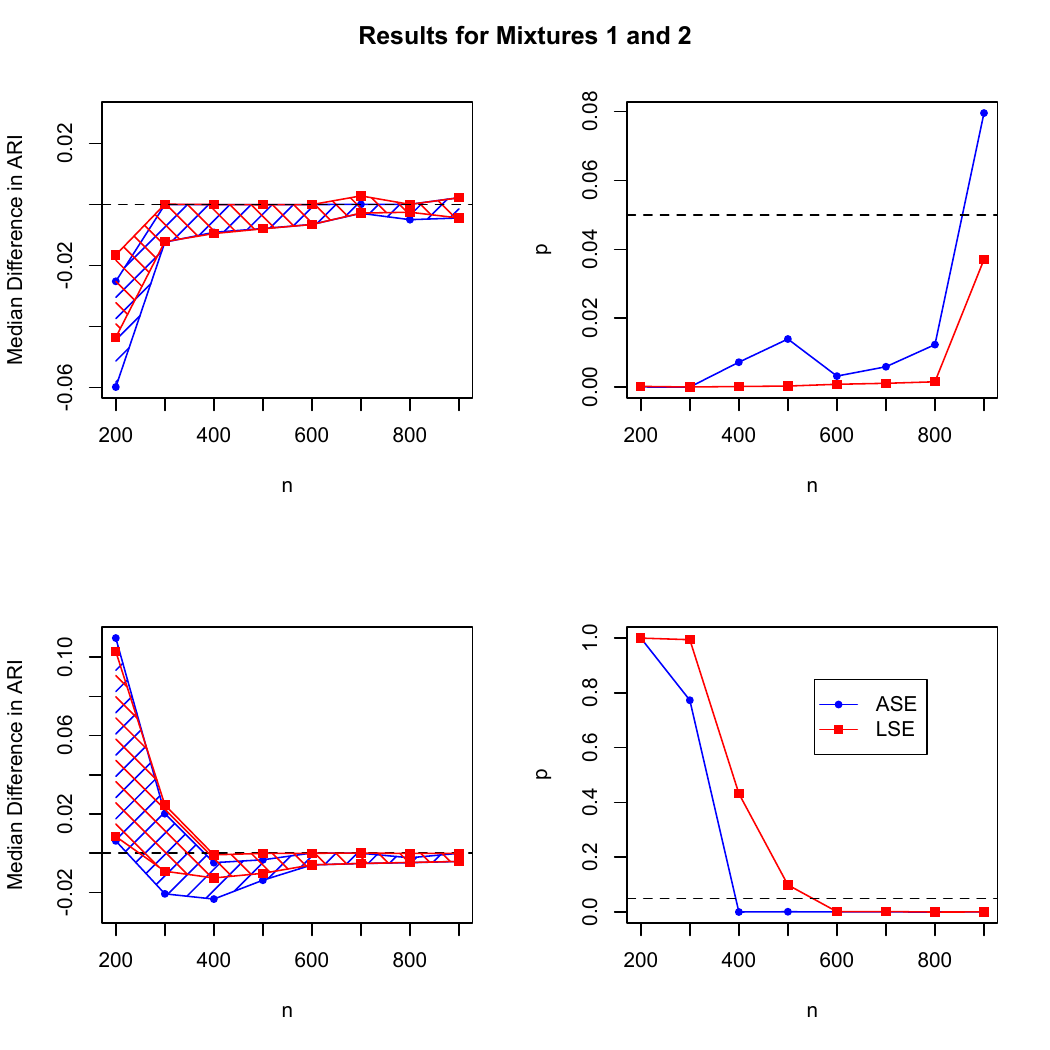}
	\caption{\label{fig:mix1_2_plot}Plots in the upper and lower rows correspond to models 1 and 2, respectively, in Section \ref{non_graph}. The left-hand plots depict 95\% confidence bands based on WIlcoxon rank-sum tests; median ARI differences below the dashed line (marking a difference of 0) indicate that ES tended to cluster more accurately than EM. The right hand plots depict $p_A$ and $p_L$, with the dashed line representing $\alpha=0.05$.}
	\end{figure}
\begin{figure}[h!]
	\centering
	\includegraphics[width = \textwidth]{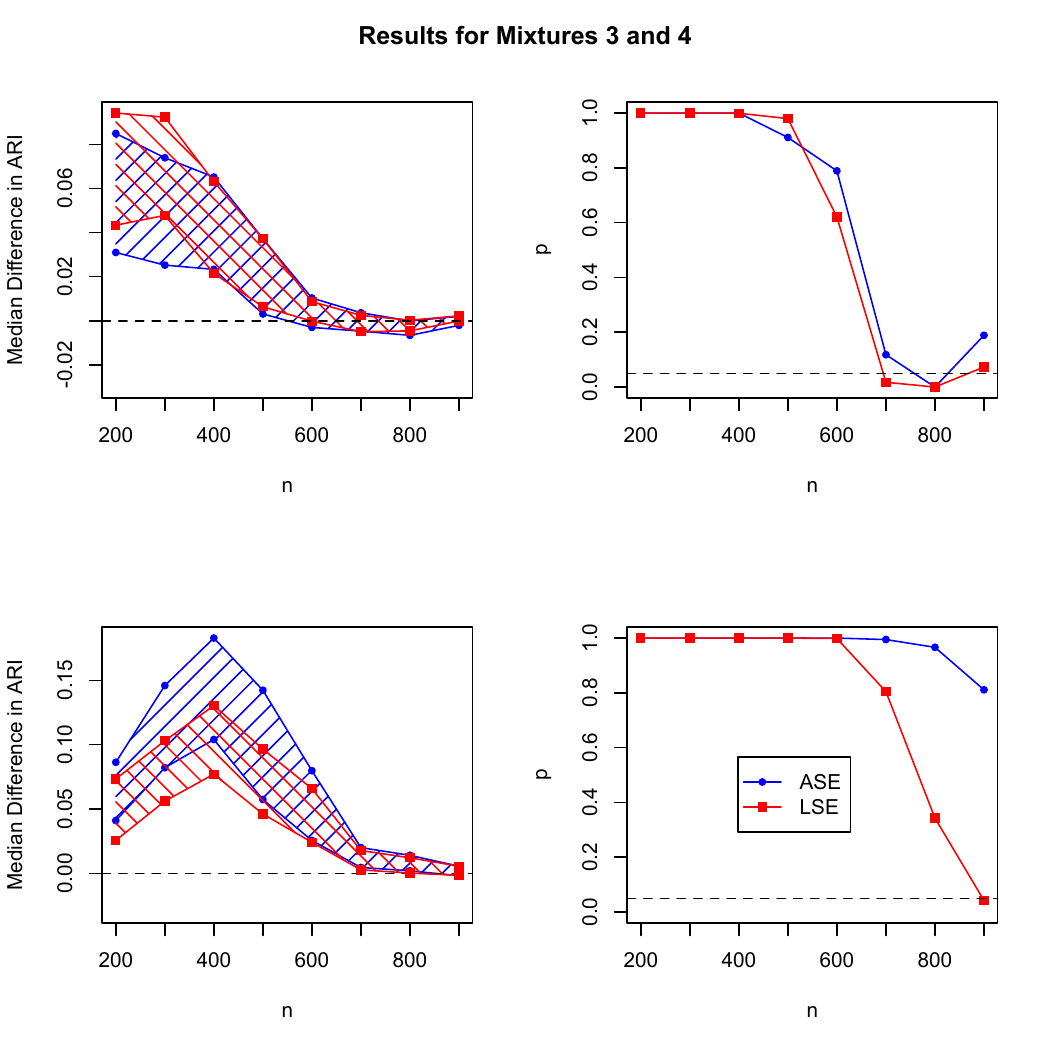}
	\caption{\label{fig:mix3_4_plot}Plots in the upper and lower rows correspond to models 3 and 4, respectively, in Section \ref{non_graph}. The plot may be interpreted analogously to Figure \ref{fig:mix1_2_plot}.}
\end{figure}
\FloatBarrier

For models 1 and 2 (Figure \ref{fig:mix1_2_plot}) the EM algorithms approximately match the ES algorithms' performance for large $n$. Due to the sample size $n$ featuring in the denominator of the component variances, the clusters of both mixtures shrink around the component means as $n$ increases, pulling the points into tight, distinct clusters.  EM tended to cluster more accurately than ES in models 3 and 4, and we credit this to the fact that in these settings the point clouds overlap significantly for small to moderate $n$, and it is not until around $n = 700$ that distinct clusters start to form.

In model 1 the ES algorithms tend to more accurately estimate the component means, variances, and weights than do the EM algorithms. As $n$ increases the accuracy of the EM algorithms begins to match that of the ES algorithms; again, we suspect that this is due to the emergence of distinct clusters. In model 2, the EM algorithms perform more accurate estimation for small $n$, but are overtaken by the ES algorithms as $n$ increases. Owing to the fact that $\textbf{x}_2$ arises as the latent positions for a sparse SBM (relative to that corresponding to $\textbf{x}_1$), we suspect that as $n$ increases beyond 900 parameter estimation by either algorithm will be comparable as this model's clusters further separate.

In models 3 and 4 (Figure \ref{fig:mix3_4_plot}) the ES algorithms tend to cluster the points less accurately than their EM counterparts, particularly so when $n$ is relatively small. Moreover, in these settings the ES algorithms largely fail to estimate the parameters more accurately than the EM algorithms, save for around $n = 800$ in model 3. Due to the decreasing trend in the confidence bounds as $n$ increases, as well as the fact that $\textbf{x}_4$ corresponds to a sparser SBM than does $\textbf{x}_3$, we suspect that this phenomenon is replicated at a range of $n > 900$ where the ES algorithms outperform the EM algorithms in model 4 as indicated by the downward trend of the $p$-values in the bottom-right plot.

\subsection{Balanced Affinity Network Structure}\label{balanced_affinity}
A $K$-block SBM is said to possess \textit{homogeneous balanced affinity structure} if $B_{ii} = a$ for all $i$, $B_{ij}=b$ for all $i\neq j$, $0<b<a<1$, and $\pi_k = \frac 1K$ for all $k$ \cite{Cape-19}. We generated 100 graphs of size $n=200, \ 500$ for each $(a,b)\in \{0.01, 0.02, \dots,0.99\}^2$ satisfying $b<a$ and compared the ARI performance of each clustering method via upper-tailed classical sign tests ($\alpha = 0.025$) due to the presence of multiple ties in several settings, with the null hypotheses:
\begin{align*}
	&H_o: P(ARI_{ES\circ ASE}\geq ARI_{EM\circ ASE}) = 0.5\\
	&H_o: P(ARI_{ES\circ LSE}\geq ARI_{EM\circ LSE}) = 0.5\\
	&H_o: P(ARI_{ES\circ ASE}\geq ARI_{ES\circ LSE}) = 0.5\\
	&H_o: P(ARI_{ES\circ LSE}\geq ARI_{ES\circ ASE}) = 0.5.
\end{align*}
The use of sharp inequalities in the proportions in question was motivated by the fact that multiple choices of $(a,b)$ result in mixtures in which perfect recovery of the cluster labels is possible by both ES and EM algorithms; moreover, if an ES algorithm manages the same ARI as its corresponding EM algorithm it does so with lower computational complexity as noted in Remark 3.

Since the latent position matrix $\textbf x$ is only identifiable up to orthogonal transformation, for each choice of $(a,b)$ we took the canonical latent position matrix to be that centered in the first quadrant of $\mathbbm R^2$ via the transformation
\begin{equation*}
	\textbf x = \boldsymbol{U_B D_B}^{\frac 12}\boldsymbol{U_B}^\top.
\end{equation*}
For each model we used the R package \textit{igraph} to sample from the SBM with the class assignments fixed, computed the ASE $\textbf{\^ X}$ from its definition, then centered the rows over the canonical latent positions with the $d\times d$ orthogonal transformation $\boldsymbol{\hat W}$ that solved the Procrustes problem
\begin{align*}
\text{min}_{\textbf{W}}\| \textbf{\^ XW}-\textbf{X}\|_F\\
\text{subject to } \textbf{W}^\top\textbf{W} = \textbf I
\end{align*}
where
\begin{equation}
\textbf{X} = \begin{bmatrix}
\boldsymbol 1_{n_1} \otimes \boldsymbol\nu_1^\top \\
\boldsymbol 1_{n_2} \otimes \boldsymbol\nu_2^\top \\
\vdots \\
\boldsymbol 1_{n_K} \otimes \boldsymbol\nu_K^\top
\end{bmatrix},
\end{equation}
with  $n_k := | \lbrace i\leq n | \tau_i = k\rbrace |$. Following this the LSE $\textbf{\v X}$ was computed via the 1--1 relationship (\ref{1-1}).

\begin{figure}[h!]
	\centering
	\includegraphics[width = \textwidth]{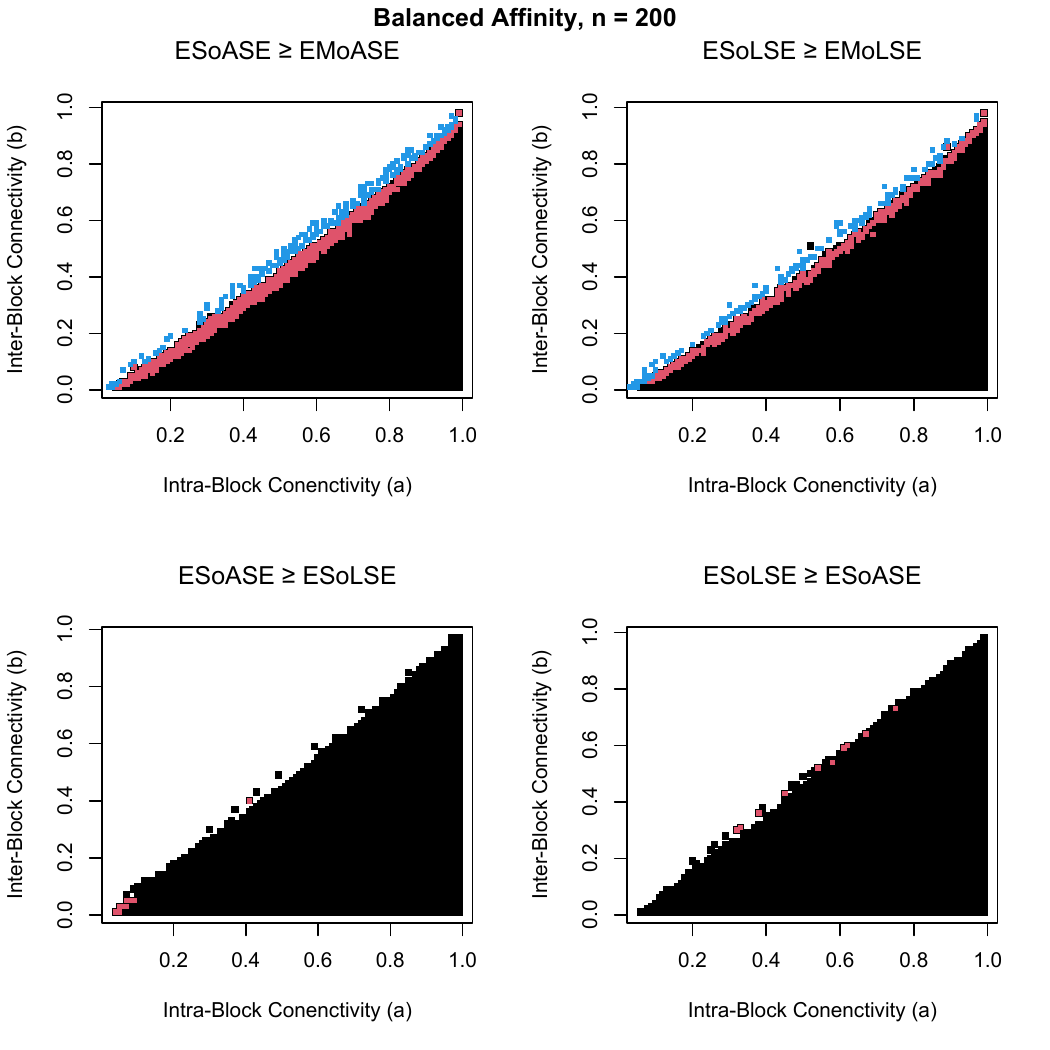}
	\caption{\label{fig:ba_plot200}Black regions denote settings in which the corresponding sharp sign test in Subsection \ref{balanced_affinity} was significant, meanwhile red regions denote significance of sign tests in which the inequality is strict. The blue regions in the upper row of plots within each figure indicate where EM strictly out-performed the corresponding ES algorithm as determined by analogous lower-tailed sign tests. have been slightly enlarged for emphasis and ease of reading. All sign tests were conducted at level $\alpha = 0.025$. Figures \ref{fig:ba_plot500}, \ref{fig:cp_plot200}--\ref{fig:cp_plot500}, and \ref{fig:cp_plot200small}--\ref{fig:cp_plot200large} may be interpreted similarly.}
\end{figure}

\begin{figure}[h!]
	\centering
	\includegraphics[width = \textwidth]{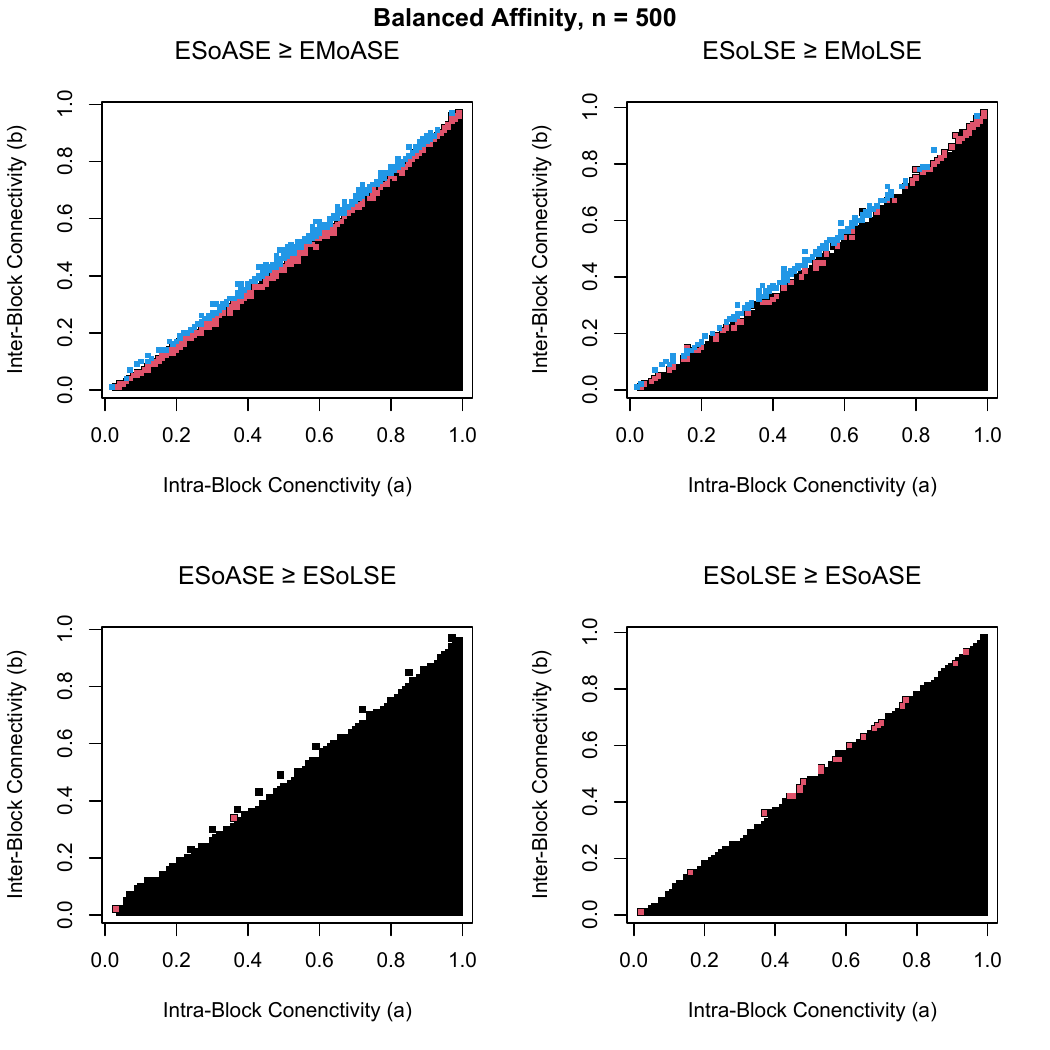}
	\caption{\label{fig:ba_plot500}}
\end{figure}

The results are displayed in Figures \ref{fig:ba_plot200}--\ref{fig:ba_plot500}. Both here and in subsequent simulations, EM tends to dominate when $(a,b)$ lies close to the identity line $a=b$ (i.e., when the simulated model is almost an Erd\"os-R\'enyi random graph \cite{JMLR}); but ES tends to dominate as the model moves away from the identity line until the true latent positions are so far apart that all methods tend to have equal performance. The red ``strictly dominant" regions shrink as $n$ increases from 200 to 500 due to the clusters shrinking around the component means, leading to larger black regions where the ES algorithms cluster at least as well as EM but with the benefit of lower computational complexity at each iteration.

\FloatBarrier

\begin{figure}[h!]
	\centering
	\includegraphics[width = \textwidth]{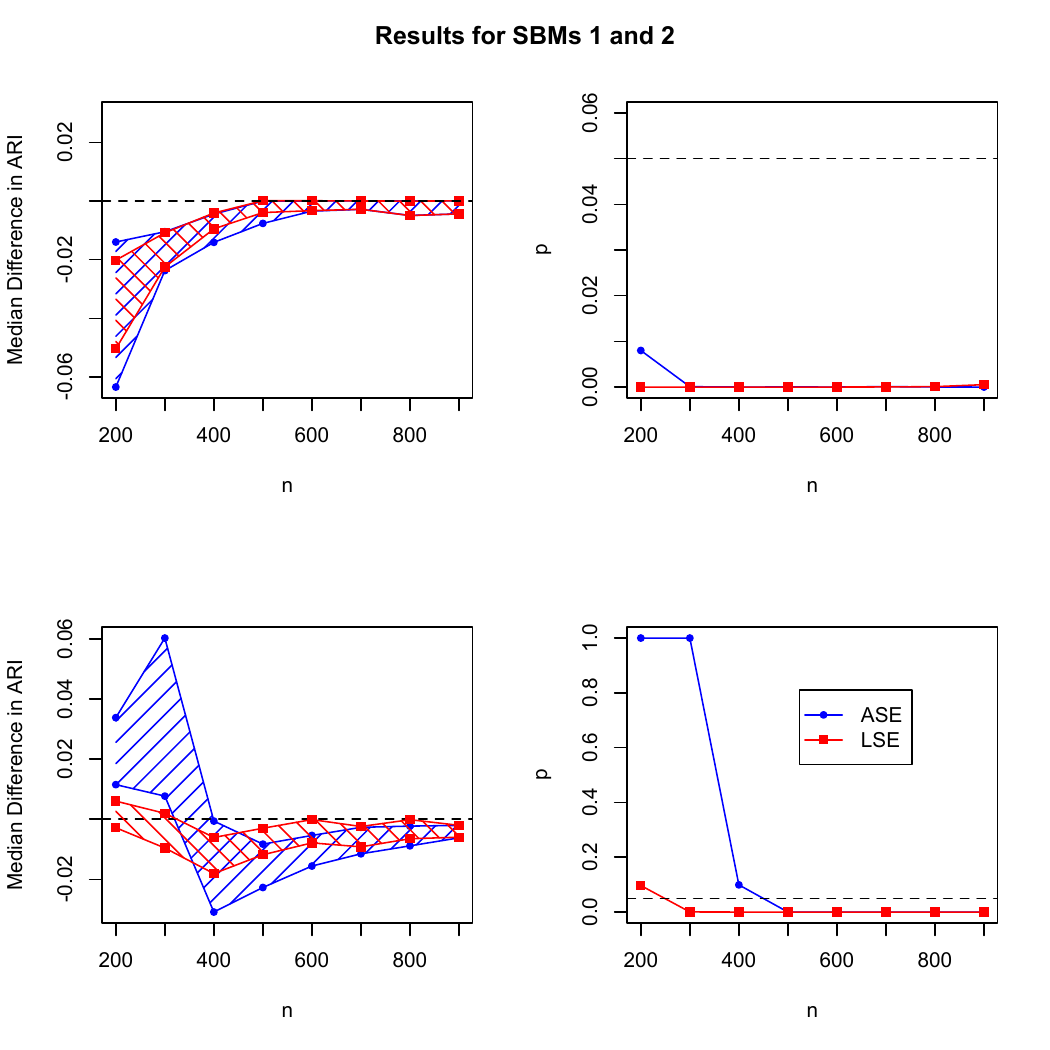}
	\caption{\label{fig:sbm1_2_plot}Plots in the upper and lower rows correspond to models 1 and 2, respectively, in Section \ref{balanced_affinity}, and may thus be interpreted analogously to Figures \ref{fig:mix1_2_plot}--\ref{fig:mix3_4_plot}.}
\end{figure}

We also repeated the simulations on $\textbf{x}_1$ and $\textbf{x}_2$ from the previous subsection, this time generating the data from balanced affinity SBMs defined by $(a,b) = (0.5, 0.4)$ (model 1) and $(a,b) = (0.2, 0.15)$ (model 2) on graphs of size $n = 200, 300, \dots, 900$. Overall we observe in Figure \ref{fig:sbm1_2_plot} similar results to those presented in Figure \ref{fig:mix1_2_plot}; however, it's clear that the degree to which ES out-clusters EM has increased, particularly in model 2.

\FloatBarrier

\subsection{Core-Periphery Network Structure}\label{balanced_core_per}
\begin{figure}[h!]
	\centering
	\includegraphics[width = \textwidth]{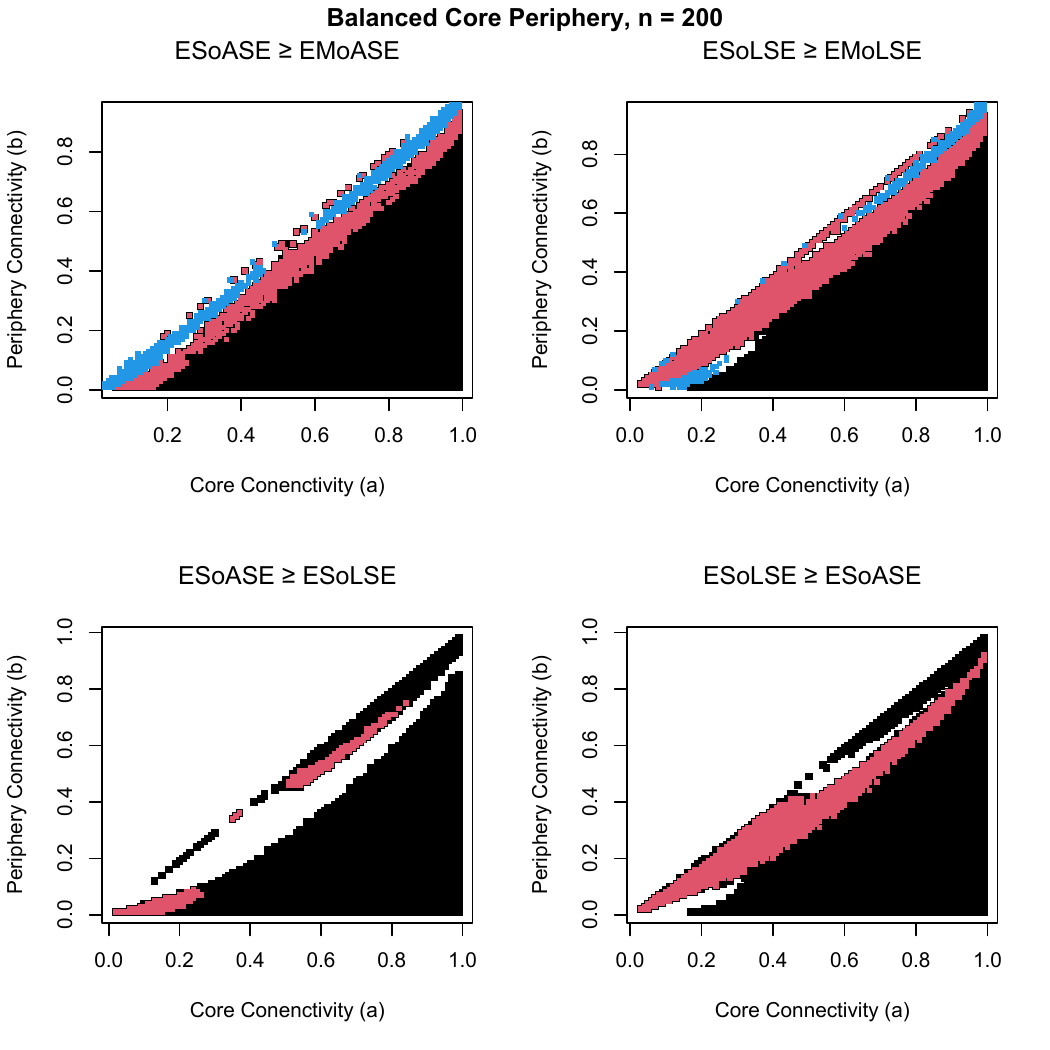}
	\caption{\label{fig:cp_plot200}}
\end{figure}

\begin{figure}[h!]
	\centering
	\includegraphics[width = \textwidth]{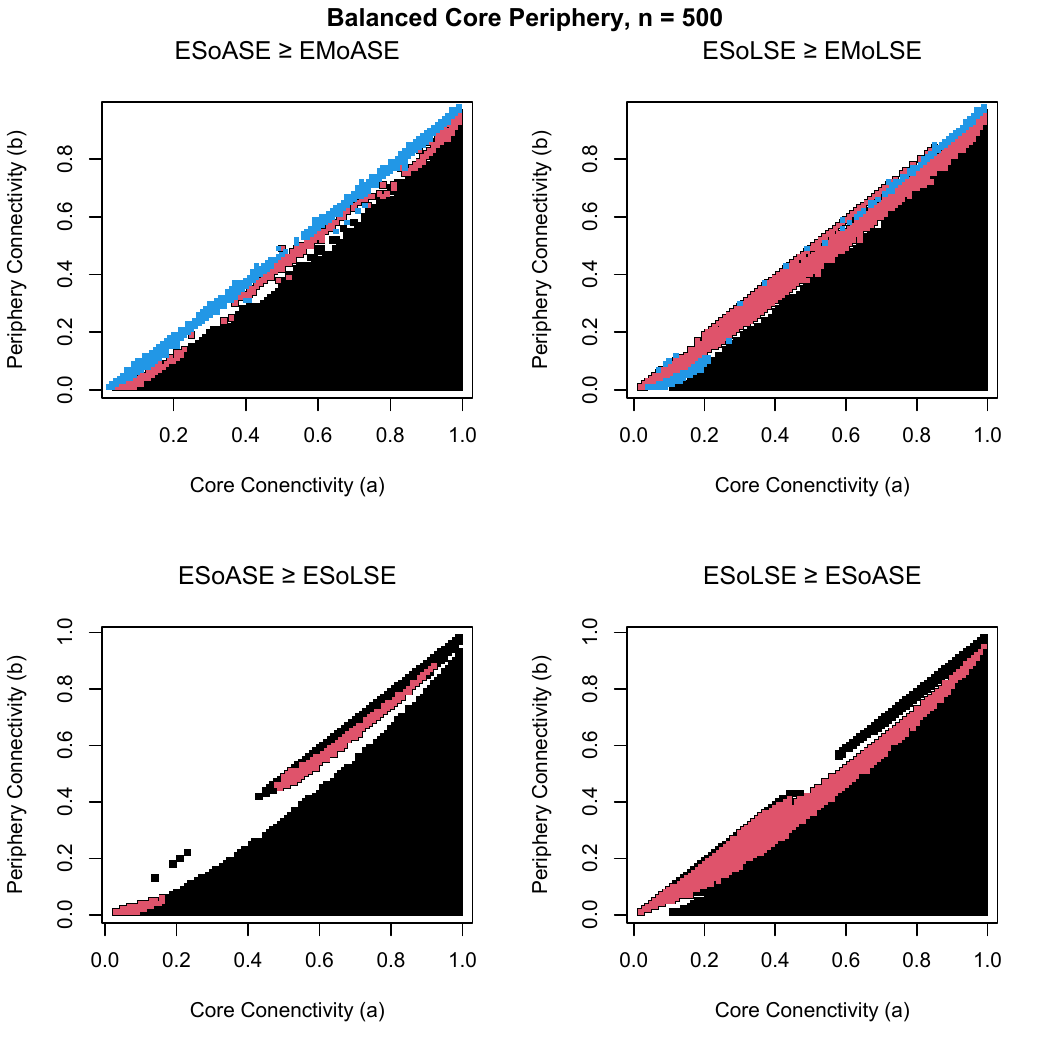}
	\caption{\label{fig:cp_plot500}}
\end{figure}

A $2$-block SBM is said to possess \textit{core-periphery structure} if $\textbf{B} = \begin{tiny}\begin{bmatrix} a &b \\ b &b\end{bmatrix}\end{tiny}$ and $\boldsymbol\pi = (\pi_1, 1-\pi_1)$ \cite{Cape-19}. We considered the 2-block balanced core-periphery SBMs characterized by the same grid of $(a,b)$ pairs in the balanced affinity simulations and repeated the previous experiment. The results are displayed in Figures \ref{fig:cp_plot200}--\ref{fig:cp_plot500}. As in the balanced affinity results we observe that $EM\circ ASE$ tends to outperform $ES\circ ASE$ near the identity line, but the latter tends dominate as $(a,b)$ moves further away from equal entries before the component means move sufficiently far apart and $ES\circ ASE$ performs at least as well as $EM\circ ASE$ with lower computational complexity. In this setting, $ES\circ LSE$ tends to dominate $EM\circ LSE$ near the identity line except for exceptionally dense models. The figures also illustrate the regions in which $ES\circ ASE$ and $ES\circ LSE$ strictly dominate each other. We  see that as $n$ increases the regions of strict dominance shrink, once again, due to the clusters shrinking around the component means.

\FloatBarrier

\begin{figure}[h!]
	\centering
	\includegraphics[width = \textwidth]{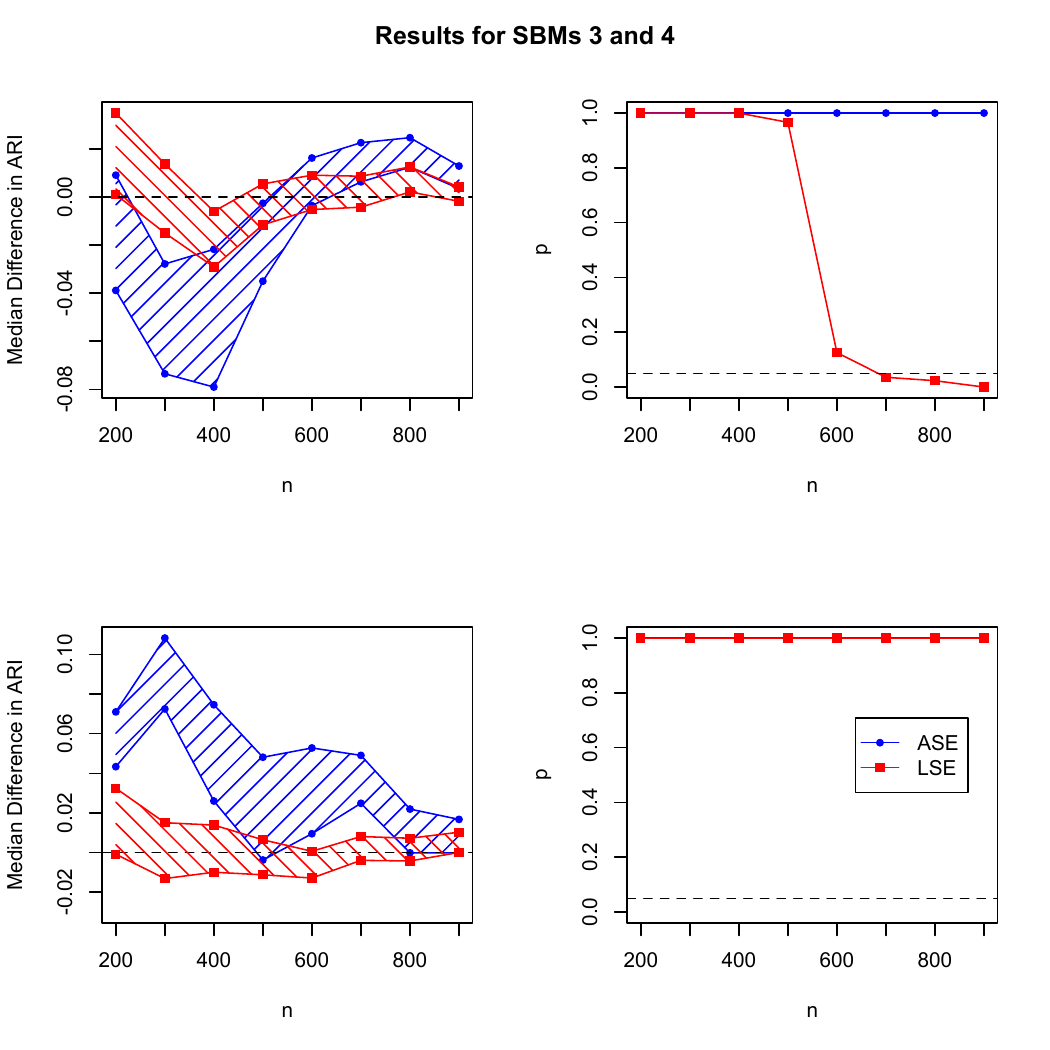}
	\caption{\label{fig:sbm3_4_plot}Plots in the upper and lower rows correspond to models 3 and 4, respectively, in Section \ref{balanced_core_per}, and may thus be interpreted analogously to Figures \ref{fig:mix1_2_plot}--\ref{fig:mix3_4_plot}, \ref{fig:sbm1_2_plot}.}
\end{figure}

We also repeated the simulations from Section \ref{non_graph} corresponding to $\textbf{x}_3$ and $\textbf{x}_4$,  which respectively arise as the latent position matrices of balanced core-periphery models with $(a,b) = (0.5,0.42)$ and $(a, b) = (0.2, 0.15)$. The results displayed in Figure \ref{fig:sbm3_4_plot} indicate that, as in the mixture setting, ES largely fails to outperform EM, except for small to moderate $n$ in model 3. However, the CIs in Figure \ref{fig:mix3_4_plot} indicate that EM tended to strictly outperform ES for both embeddings; but in Figure \ref{fig:sbm3_4_plot} , $ES \circ LSE$ is approximately on par with $EM\circ LSE$. Both here and in the mixture setting, it was observed that ES tended to vastly overestimate the entries of the covariance matrices, hence the algorithms' seeming inability to more accurately estimate $\boldsymbol \Psi$, except for $ES\circ LSE$ in model 3 when $700 \leq n \leq 900$.

\FloatBarrier

\subsection{Unbalanced Core-Periphery Model}

We have to this point only examined the performance of ES against that of EM for balanced 2-block SBMs, in which the blocks have equal probability of node membership. Since this case obviously fails to encapsulate a wide variety of models, we also evaluated the performance of the clustering procedures for unbalanced core-periphery model in which the more densely connected core possesses a membership probability of either 0.25 (``Small Core'') and 0.75 (``Large Core'') in graphs of size $n = 200$.

\begin{figure}[h!]
	\centering
	\includegraphics[width = \textwidth]{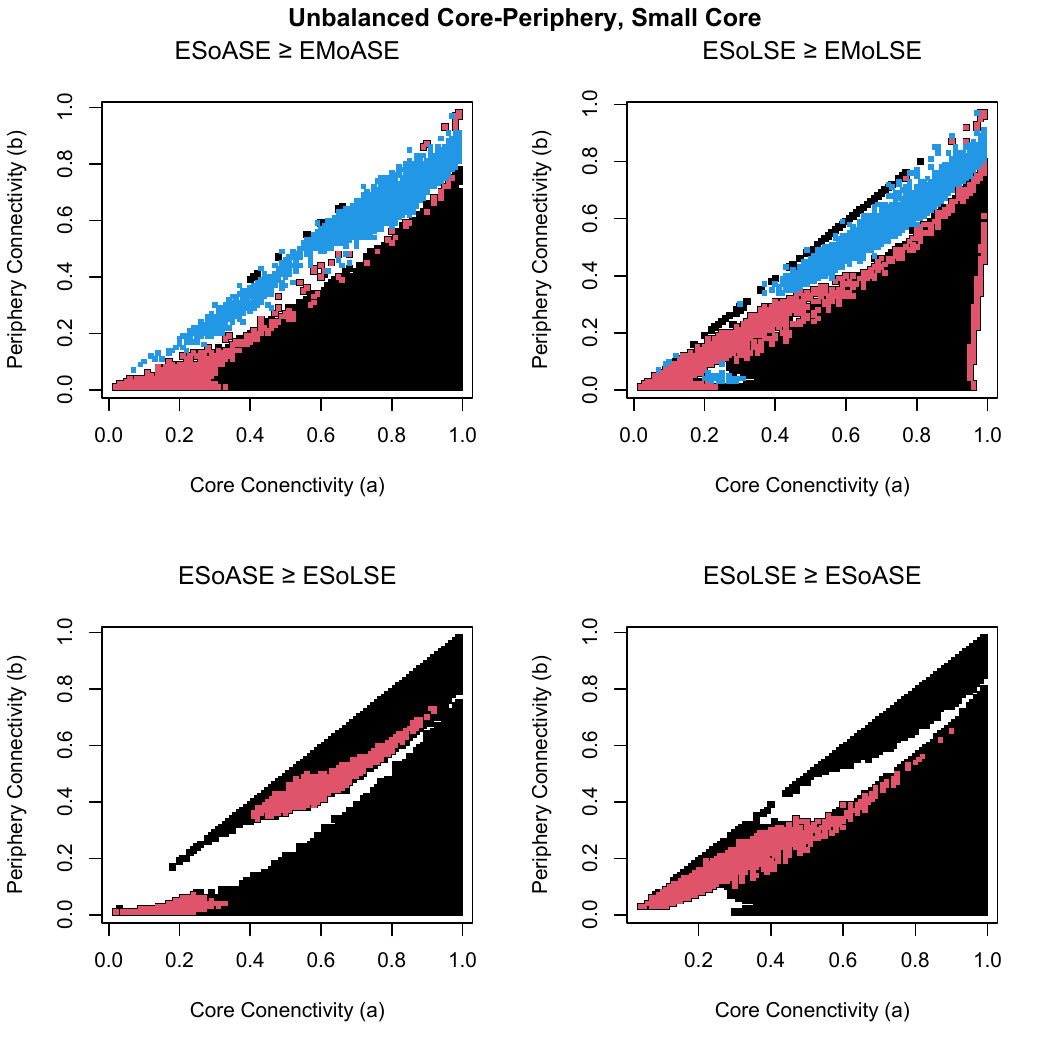}
	\caption{\label{fig:cp_plot200small}}
\end{figure}

\begin{figure}[h!]
	\centering
	\includegraphics[width = \textwidth]{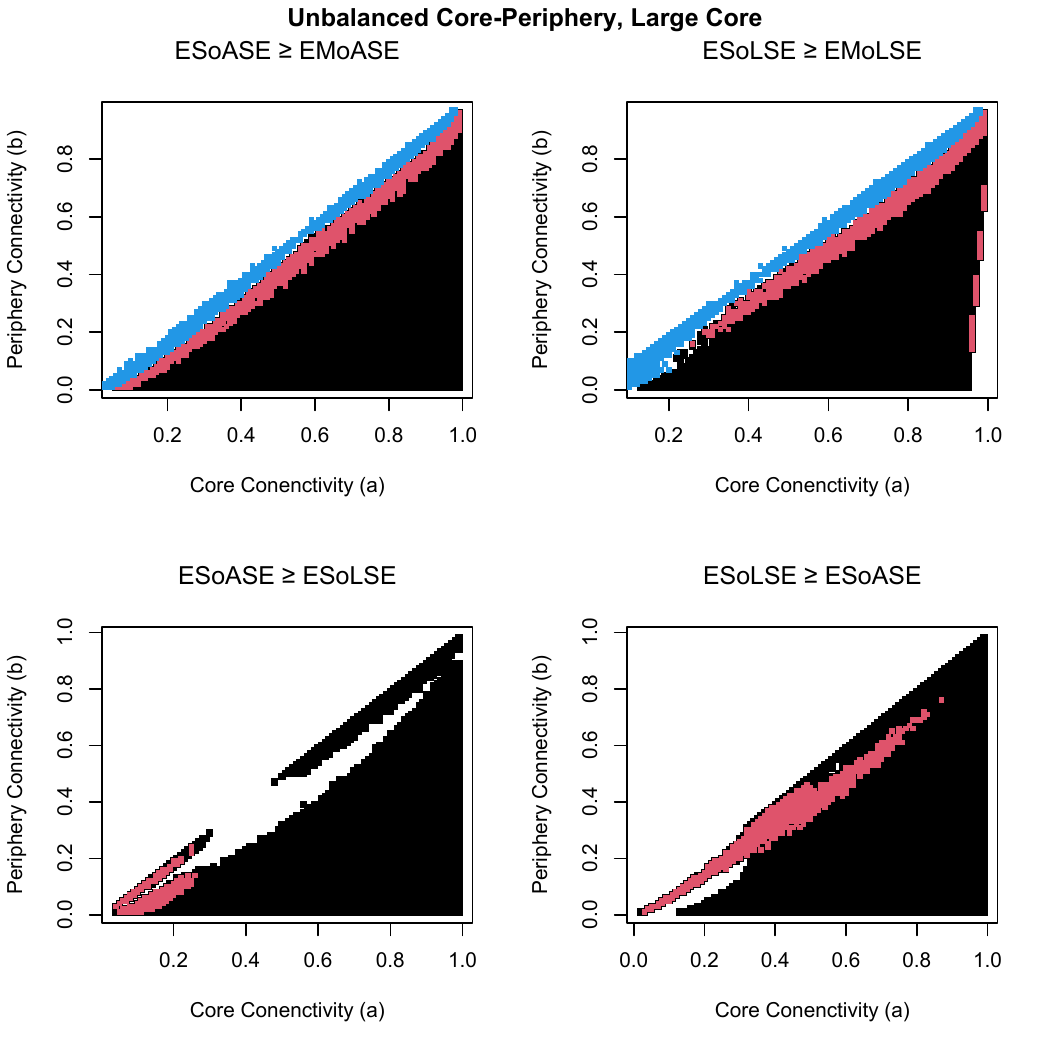}
	\caption{\label{fig:cp_plot200large}}
\end{figure}

The results are presented in Figures \ref{fig:cp_plot200small}--\ref{fig:cp_plot200large}. We see that the ES algorithms tend to dominate EM in sparse settings when the core is small, but when the core is large EM tends to dominate in a larger region near the diagonal than in balanced settings. These plots along with Figure \ref{fig:cp_plot200} also illustrate how regions in which the ES algorithms strictly dominate each other shift as the core proportion changes.

\FloatBarrier

\subsection{Synthetic Analysis of MRI Connectome SBM}
\cite{Priebe-2019} investigated clustering via the EM algorithm for the ASE and LSE of a 4-block SBM used to model neural connectivity in the human brain. They noted that the 2-dimensional LSE captures left-hemisphere/right-hemisphere connectivity and the 2-dimensional ASE captures gray-matter/white-matter connectivity. That is, clustering according to either embedding resulted in one of ``two truths.''

To compare clustering performance for this setting, we generated data from the 4-block connectome estimated by \cite{Priebe-2019}, i.e., the model $\textbf{A} \sim \text{SBM}(\textbf{B}, \boldsymbol\pi)$ where $\boldsymbol\pi = (0.28,0.22,0.28,0.22)$, $\textbf{B}$ is the rank-4 matrix
\begin{equation*}
\textbf{B} = \begin{bmatrix}
0.020 &0.044 &0.002 &0.009\\
0.044 &0.115 &0.010 &0.042\\
0.002 &0.010 &0.020 &0.045\\
0.009 &0.042 &0.045 &0.117
\end{bmatrix},
\end{equation*}
and blocks 1--4 correspond to left/gray, left/white, right/gray, and right/white neurons, respectively. The eigendecomposition $\textbf{B} = \textbf{U}_{\textbf{B}} \textbf{D}_{\textbf{B}} \textbf{U}_{\textbf{B}}^\top $ gives the latent position matrix (after rotating properly to the first orthant in $\mathbbm R^4$) as
\begin{equation*}
	\textbf x = \textbf{B} = \textbf{U}_{\textbf{B}} \textbf{D}_{\textbf{B}}^{\frac12} \textbf{U}_{\textbf{B}}^\top = 
	\begin{bmatrix}
	\boldsymbol\nu_1^\top \\
	\boldsymbol\nu_2^\top \\
	\boldsymbol\nu_3^\top \\
	\boldsymbol\nu_4^\top
	\end{bmatrix}
	=
	\begin{bmatrix}
	0.0915 &\ 0.1076 &\ 0.0057 &\ 0.0034 \\
	0.1076 &\ 0.3149 &\ 0.0056 &\ 0.0649 \\
	0.0057 &\ 0.0056 &\ 0.0886 &\ 0.1099 \\
	0.0034 &\ 0.0649 &\ 0.1099 &\ 0.3173
	\end{bmatrix}.
\end{equation*}

\begin{figure}[h!]
	\centering
	\includegraphics[width = \textwidth]{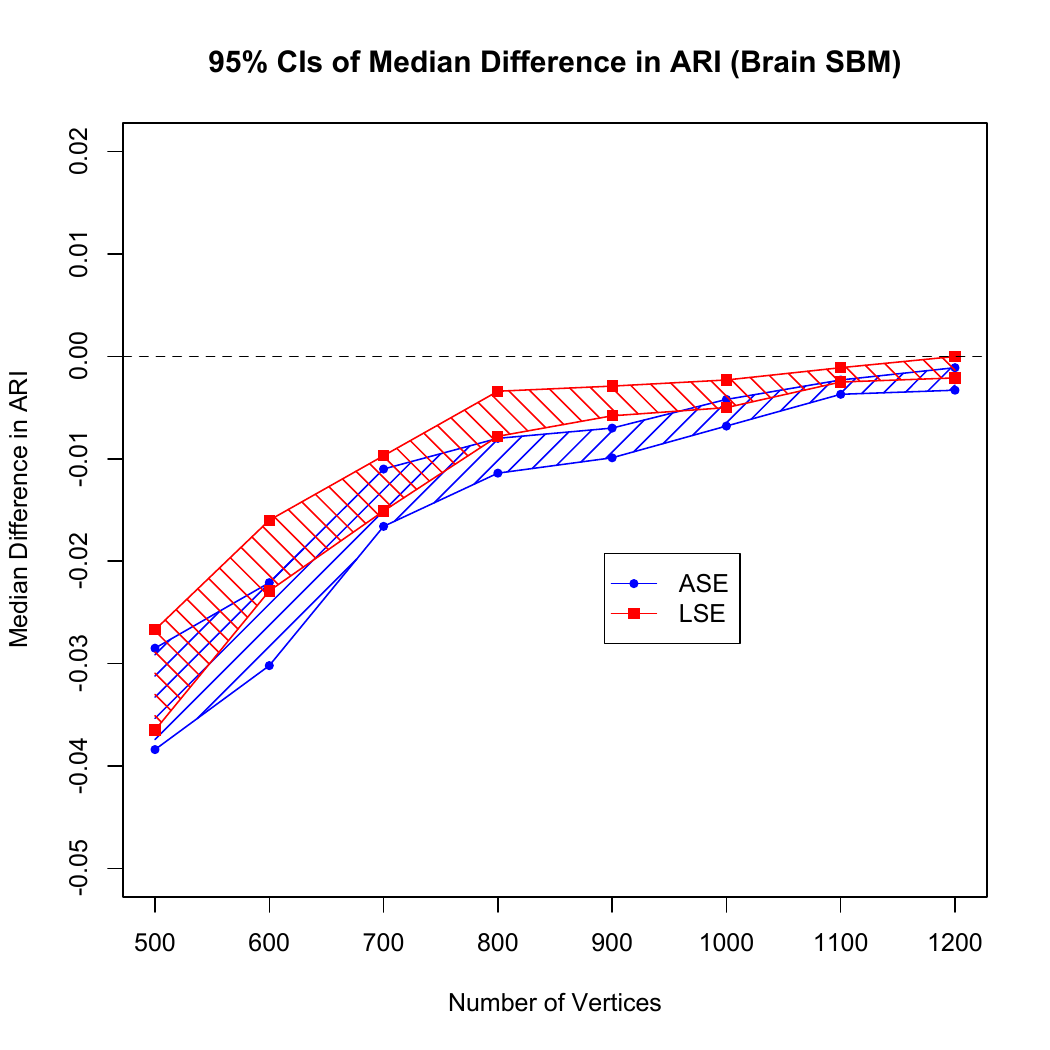}
	\caption{\label{fig:brain_sbm_results_plot}}
\end{figure}

For each $n = 500, 600, \dots, 1200$ we generated 100 graphs, computed their ASEs and LSEs, carried out the procedures as otherwise described, and output the results to Figure \ref{fig:brain_sbm_results_plot}. We opted not to compare accuracy of parameter estimation, since we observed that our algorithms tended to vastly overestimate the covariances as described in the previous subsection. Here ES tended to more successfully cluster than EM for all values of $n$.


\FloatBarrier

\subsection{Rank-Deficient SBM}\label{low_rank_sims}

\begin{figure}[h!]
	\centering
	\includegraphics[width = \textwidth]{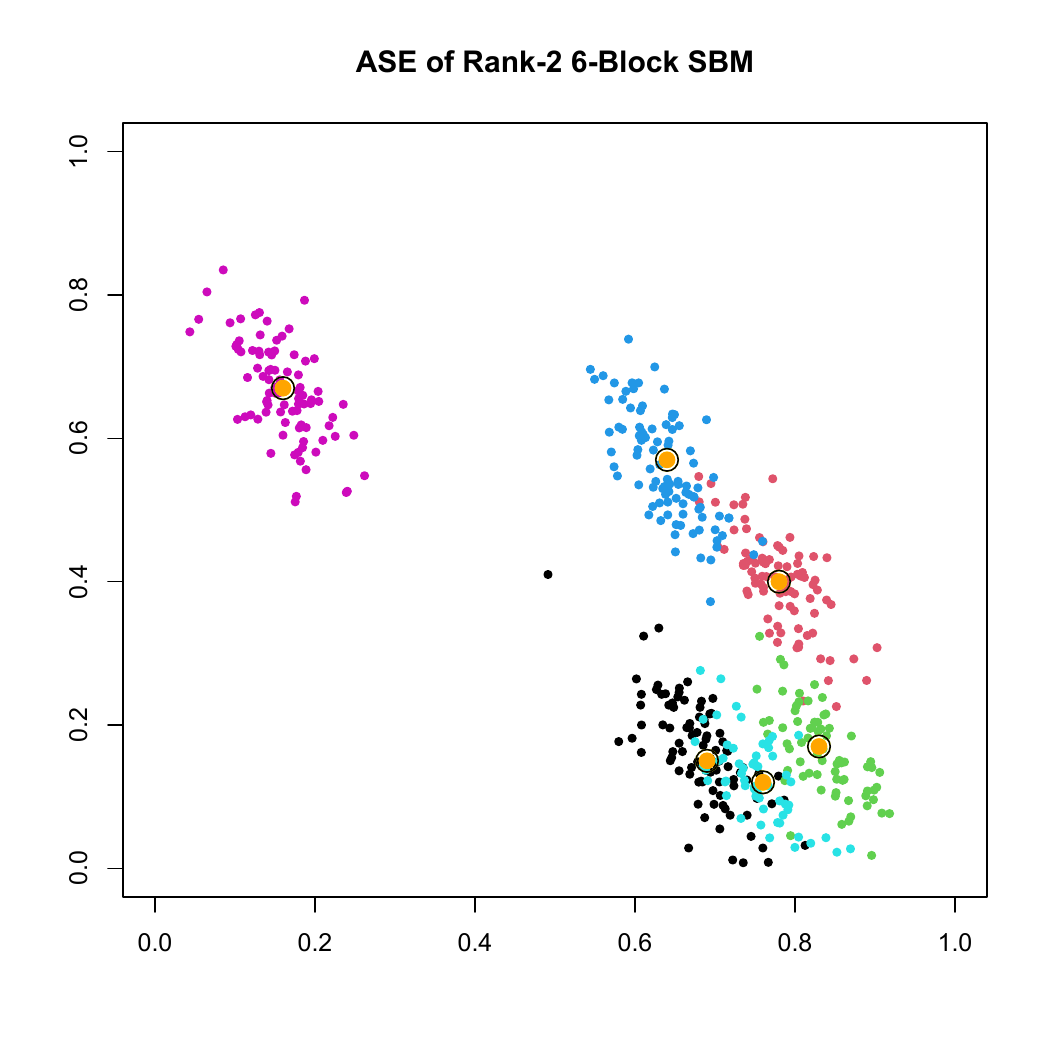}
	\caption{\label{fig:low_rank_ase_plot}The smaller points representing rows of the 2-dimensional ASE have been colored according to block membership of the corresponding nodes in the original graph. The larger orange points encircled with a black outline denote the true latent positions, i.e., the rows of $\textbf{X}$.}
\end{figure}

\begin{figure}[h!]
	\centering
	\includegraphics[width = \textwidth]{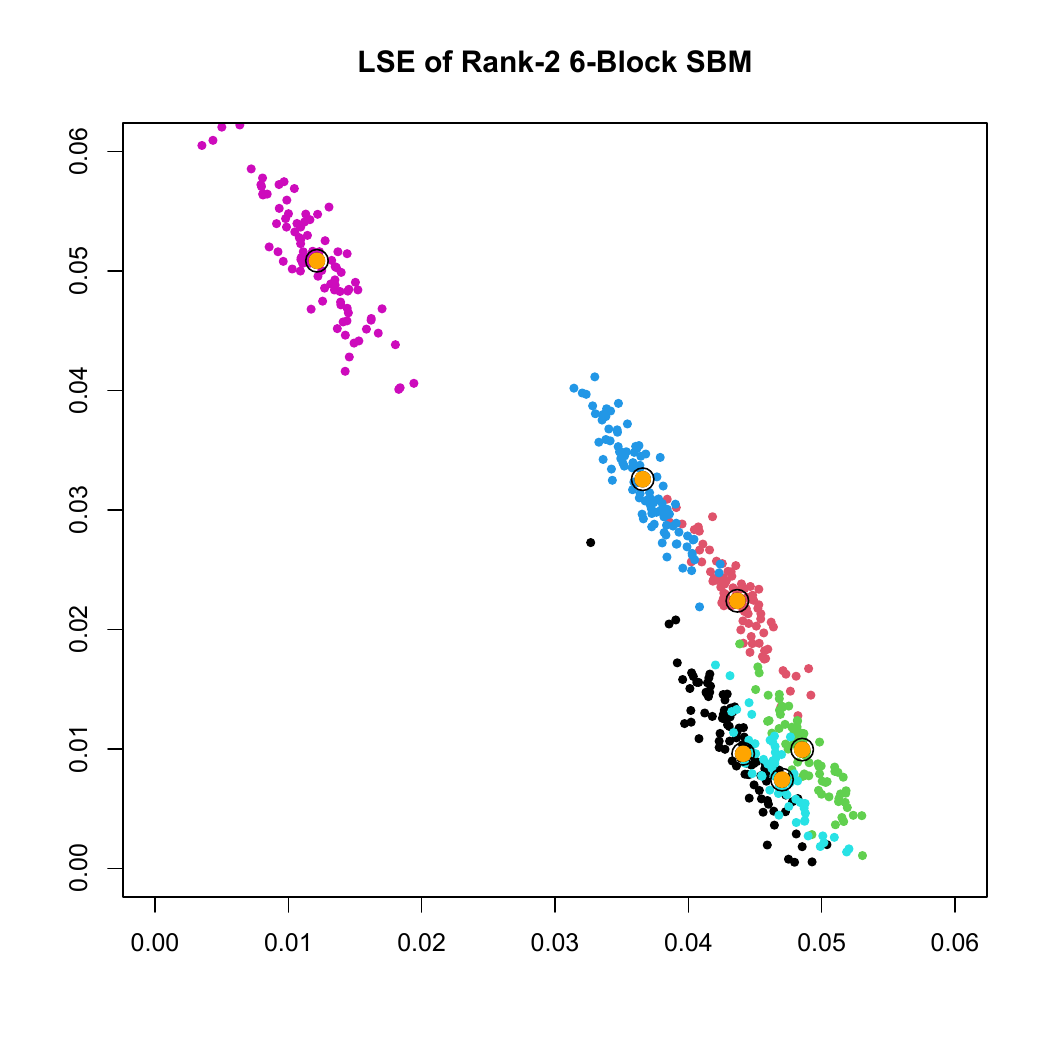}
	\caption{\label{fig:low_rank_lse_plot}The smaller points representing rows of the 2-dimensional LSE have been colored according to block membership of the corresponding nodes in the original graph (as in the previous figure). Here the larger orange points denote the scaled Laplacian latent positions, i.e., the component means of the LSE's limiting distribution.}
\end{figure}

All of the settings in which we have conducted our simulations thus far share the property of the true rank $d$ being equal to the number of blocks $K$; of course, one may quite naturally give an example of SBM in which $d < K$. To this end, consider the 6-block SBM given by
\begin{equation*}
	\textbf{B} = \begin{bmatrix}
		0.4986 &\ 0.5982 &\ 0.5982 &\ 0.5271 &\ 0.5424 &\ 0.2109\\
		0.5982 &\ 0.7684 &\ 0.7154 &\ 0.7272 &\ 0.6408 &\ 0.3928\\
		0.5982 &\ 0.7154 &\ 0.7178 &\ 0.6281 &\ 0.6512 &\ 0.2467\\
		0.5271 &\ 0.7272 &\ 0.6281 &\ 0.7345 &\ 0.5548 &\ 0.4843\\
		0.5424 &\ 0.6408 &\ 0.6512 &\ 0.5548 &\ 0.5920 &\ 0.2020\\
		0.2109 &\ 0.3982 &\ 0.2467 &\ 0.4843 &\ 0.2020 &\ 0.4745
	\end{bmatrix}
\end{equation*}
and $\boldsymbol\pi = (0.19, 0.18, 0.14, 0.19 ,0.12, 0.18)$. The block probability matrix is rank-2, and may be exactly rewritten as $\textbf{B} = \textbf{xx}^\top$, where
\begin{equation*}
	\textbf x = \begin{bmatrix}
		0.69 &\ 0.15\\
		0.78 &\ 0.40\\
		0.83 &\ 0.17\\
		0.64 &\ 0.57\\
		0.76 &\ 0.12\\
		0.16 &\ 0.67
	\end{bmatrix}.
\end{equation*}
Unlike the above connectome SBM this particular setting was not motivated by a particular application. In fact, we randomly generated this model by sampling the six rows of $\textbf x$ uniformly from the intersection of the 2-dimensional unit disc and the positive quadrant in $\mathbbm R^2$ and then generating $\boldsymbol\pi$ from a flat Dirichlet distribution (with everything rounded to two decimal places). We have provided examples of the ASE and LSE of a graph generated from this setting $(n =600)$ in Figures \ref{fig:low_rank_ase_plot}--\ref{fig:low_rank_lse_plot}.

\begin{figure}[h!]
	\centering
	\includegraphics[width = \textwidth]{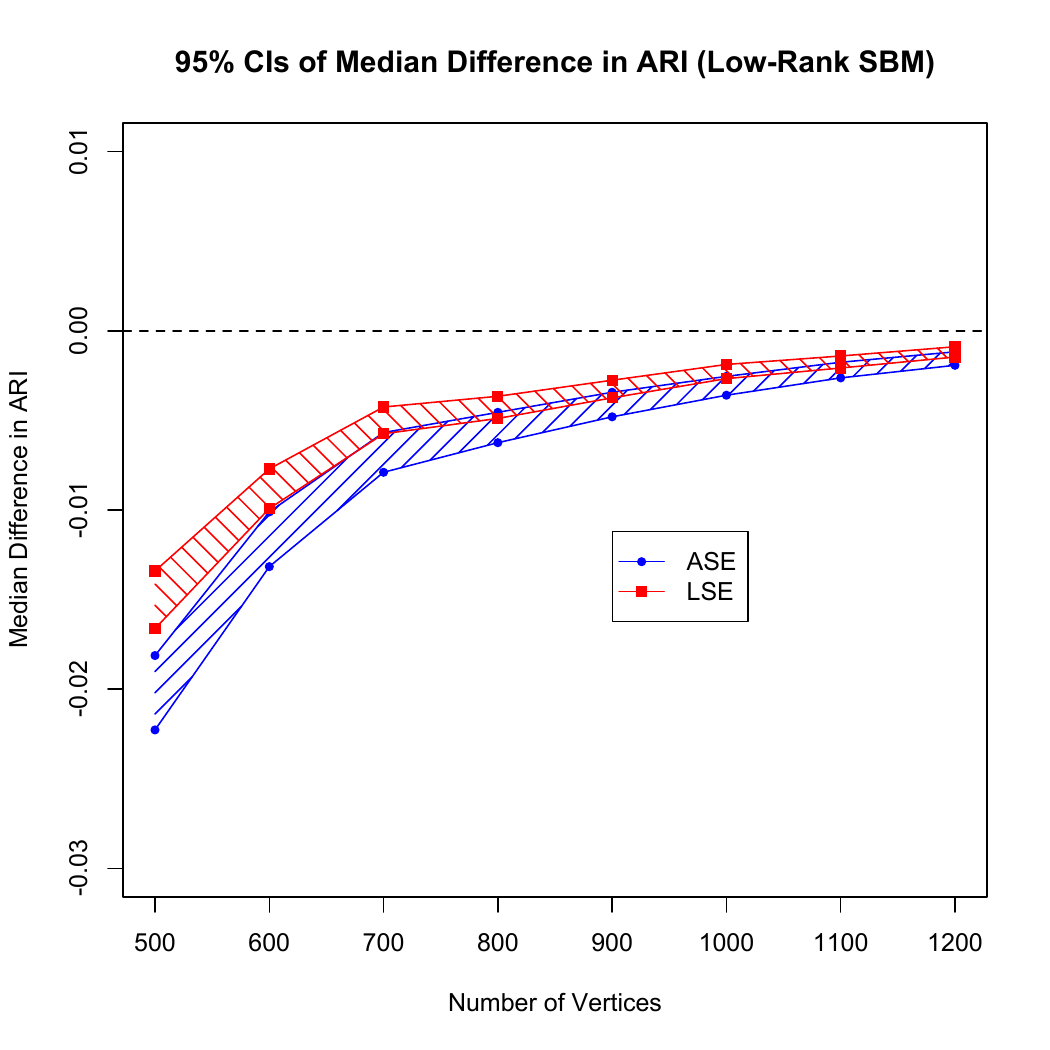}
	\caption{\label{fig:low_rank_results_plot}}
\end{figure}

We repeated the experiment of the connectome SBM, this time electing to generate 1000 graphs for each $n$, and output the results to Figure \ref{fig:low_rank_results_plot}. One can see that both ES algorithms dominate their corresponding EM for this particular setting. We include the results of additional rank-deficient simulations in Appendix \ref{app:rank_deficient_appendix}.

\FloatBarrier

\section{Discussion}\label{discussion}

We have described an algorithm that estimates the parameters of a curved-normal mixture model and accounts for the components' curvature. The algorithm itself is an adaptation of the usual EM algorithm for smooth-normal mixture models, where instead of updating the component variance estimates in the usual way we simply plug the usual updates to the component means into the variance function(s). Even though we developed this algorithm purely for the purpose of spectral clustering for SBMs, we suspect that we can easily adapt it to mixtures of curved exponential families in which the component distributions are not normal. We hope to explore this, as well as sufficient conditions for consistent and asymptotically normal estimates as outlined in the appendix of \cite{ES-algorithm}, in a future paper.

The simulation results demonstrate that taking into account the curvature of the spectral embeddings' limiting distributions allows us to improve node clustering for SBMs, and --- in some settings --- by a vast margin. In general our results indicate that EM dominates ES in settings close to Erd\"os-R\'enyi models, but ES overtakes EM in models sufficiently different from such a simple setting. In particular, the dominance our algorithms display in outperforming EM for the brain connectome model lead us to highly recommend our method for that application. Moreover, there existed at least one sample size for which at least one of our proposed algorithms clustered significantly more accurately than vanilla EM in all but one of the four SBMs which we singled out in our Subsections \ref{balanced_affinity} and \ref{balanced_core_per}. Since we initialized all simulations at the true parameter values, we have omitted any discussion of sufficient conditions for local or global convergence of $ES\circ\{ASE, LSE\}$. However, we observed anecdotally in the brain connectome and rank-deficient settings that initializing ES far from the canonical latent positions still resulted in accurate clusters. The default implementation of $mclust$ initializes EM at the terminating values of a hierarchical agglomerative clustering procedure \cite{Raftery-16}, which may also prove practically suitable for ES in future work.

Though the simulations are presented through the lens of evaluating ES against EM, we also compared both methods to $K$-means (Appendix \ref{app:appendix2}). When all three algorithms were compared to each other, we determined that their clustering performance was approximately equal for the homogeneous balanced affinity models, and that EM and ES performed vastly better than $K$-means in the other settings. This corroborates the prior work done by \cite{Avanti-16} and \cite{Tang-18}; since the canonical latent positions of a 2-dimensional full-rank homogeneous balanced affinity model lie equally spaced in the first quadrant with equal covariances, the $K$-means assumption of spherical covariances does not particularly hinder the clustering problem. As the other settings are far more varied in the spacing of their latent positions and the shape and orientation of their covariances, the flexibility afforded by EM and ES renders them far more effective than $K$-means.

The major drawback of our algorithms in the SBM setting is their seeming inability to accurately estimate the component variances. We noted that ES occasionally yielded values of $\boldsymbol\Lambda^{-1}$ and $\boldsymbol{\tilde\Lambda}^{-1}$ with diagonal entries far exceeding 1. A possible alteration to the algorithm that may reduce or eliminate this issue entirely would be to replace $\boldsymbol\Lambda$, $\boldsymbol \mu$, and $\boldsymbol{\tilde\Lambda}$ with
\begin{align*}
	\boldsymbol{\hat\Lambda} &= \frac{\sum_{i=1}^n\textbf{\^ X}_i \textbf{\^ X}_i^\top }{n}\\
	\boldsymbol{\hat\mu} &= \frac{\sum_{i=1}^n \textbf{\^ X}_i}{n}\\
	\boldsymbol{\hat{\tilde\Lambda}} &= \frac{\sum_{i=1}^n\frac{\textbf{\^ X}_i \textbf{\^ X}_i^\top }{\textbf{\^ X}_i^\top  \boldsymbol{\hat \mu}}}{n},
\end{align*}
respectively, since doing so holds those terms constant and would prevent slight perturbations to the component means from causing excessive inflation to the entries of the component variances. Doing so also drastically reduces the computational complexity of each iteration to those mentioned in Remark \ref{rem:complexity}.

Nonetheless, the strong evidence that spectral clustering based on ES can dominate that of EM opens up avenues for future research. During the writing of this article, \cite{xie-19} developed one-step estimators for the latent position matrices of both the ASE and LSE, the rows of which also asymptotically converge to curved normal mixtures. These estimators make additional use of the likelihood structure of the underlying random graph model, and as a result the ensuing C-GMM possesses component variances that are locally more efficient than those of the mixtures we have considered thus far. Moreover, they find that EM clustering based on these estimators can improve upon $EM\circ\{ASE, LSE\}$. Further work due to \cite{xie-21} modified the limiting distributions to include the sparsity factor $\rho_n$ in the covariance functions. Moreover, all of our simulations were conducted under the assumption that the rank of the block probability matrix $d$ and number of blocks $K$ are both known; hence, we have left untreated the model selection problem of deciding upon these values from a collection of candidates. \cite{yang-21} developed an extended ASE, which also takes a limiting C-GMM, to address this problem exactly. In future articles we hope to compare performance of $ES\circ\{ASE, LSE\}$ with EM for these additional estimators, as well as implement ES algorithms based on the latter.

\bibliographystyle{imsart-nameyear.bst} 
\bibliography{ES_for_GMM}       

\begin{thebibliography}{27}

\bibitem[\protect\citeauthoryear{Akaike}{1974}]{AIC}
\begin{barticle}[author]
\bauthor{\bsnm{Akaike},~\bfnm{Hirotugu}\binits{H.}}
(\byear{1974}).
\btitle{A new look at the statistical model identification}.
\bjournal{IEEE Transactions on Automatic Control}
\bvolume{16}
\bpages{716-723}.
\end{barticle}
\endbibitem

\bibitem[\protect\citeauthoryear{Athreya et~al.}{2016}]{Avanti-16}
\begin{barticle}[author]
\bauthor{\bsnm{Athreya},~\bfnm{Avanti}\binits{A.}},
  \bauthor{\bsnm{Priebe},~\bfnm{Carey~E.}\binits{C.~E.}},
  \bauthor{\bsnm{Tang},~\bfnm{Minh}\binits{M.}},
  \bauthor{\bsnm{Marchette},~\bfnm{David~J.}\binits{D.~J.}} \AND
  \bauthor{\bsnm{Sussman},~\bfnm{Daniel~L.}\binits{D.~L.}}
(\byear{2016}).
\btitle{A limit theorem for scaled eigenvectors of random dot product graphs}.
\bjournal{Sankhya A}
\bvolume{78}
\bpages{1-18}.
\end{barticle}
\endbibitem

\bibitem[\protect\citeauthoryear{Athreya et~al.}{2018}]{JMLR}
\begin{barticle}[author]
\bauthor{\bsnm{Athreya},~\bfnm{Avanti}\binits{A.}},
  \bauthor{\bsnm{Fishkind},~\bfnm{Donniell~E.}\binits{D.~E.}},
  \bauthor{\bsnm{Tang},~\bfnm{Minh}\binits{M.}},
  \bauthor{\bsnm{Park},~\bfnm{Youngser}\binits{Y.}},
  \bauthor{\bsnm{Vogelstein},~\bfnm{Joshua~T.}\binits{J.~T.}},
  \bauthor{\bsnm{Levin},~\bfnm{Keith}\binits{K.}},
  \bauthor{\bsnm{Lyzinski},~\bfnm{Vince}\binits{V.}},
  \bauthor{\bsnm{Qin},~\bfnm{Yichen}\binits{Y.}} \AND
  \bauthor{\bsnm{Sussman},~\bfnm{Daniel~L.}\binits{D.~L.}}
(\byear{2018}).
\btitle{Statistical inference on random dot product graphs: A survey}.
\bjournal{Journal of Machine Learning Research}
\bvolume{18}
\bpages{1-92}.
\end{barticle}
\endbibitem

\bibitem[\protect\citeauthoryear{Bickel and Doksum}{2015}]{Bic-Dok}
\begin{bbook}[author]
\bauthor{\bsnm{Bickel},~\bfnm{Peter~J.}\binits{P.~J.}} \AND
  \bauthor{\bsnm{Doksum},~\bfnm{Kjell~A.}\binits{K.~A.}}
(\byear{2015}).
\btitle{Mathematical Statistics: Basic Ideas and Selected Topics}
\bvolume{1},
\bedition{2nd} ed.
\bpublisher{CRC Press}.
\end{bbook}
\endbibitem

\bibitem[\protect\citeauthoryear{Cape, Tang and Priebe}{2019}]{Cape-19}
\begin{barticle}[author]
\bauthor{\bsnm{Cape},~\bfnm{Joshua}\binits{J.}},
  \bauthor{\bsnm{Tang},~\bfnm{Minh}\binits{M.}} \AND
  \bauthor{\bsnm{Priebe},~\bfnm{Carey~E.}\binits{C.~E.}}
(\byear{2019}).
\btitle{On spectral embedding performance and elucidating network structure in
  stochastic block model graphs}.
\bjournal{Network Science}
\bvolume{7}
\bpages{269-291}.
\end{barticle}
\endbibitem

\bibitem[\protect\citeauthoryear{Dempster, Laird and Rubin}{1977}]{DLR}
\begin{barticle}[author]
\bauthor{\bsnm{Dempster},~\bfnm{Arthur~P.}\binits{A.~P.}},
  \bauthor{\bsnm{Laird},~\bfnm{Nan~M.}\binits{N.~M.}} \AND
  \bauthor{\bsnm{Rubin},~\bfnm{Donald~B.}\binits{D.~B.}}
(\byear{1977}).
\btitle{Maximum likelihood from incomplete data via the {EM} algorithm}.
\bjournal{Journal of the Royal Statistical Society. Series B (Methodological)}
\bvolume{39}
\bpages{1-38}.
\end{barticle}
\endbibitem

\bibitem[\protect\citeauthoryear{Elashoff and Ryan}{2004}]{ES-algorithm}
\begin{barticle}[author]
\bauthor{\bsnm{Elashoff},~\bfnm{Michael}\binits{M.}} \AND
  \bauthor{\bsnm{Ryan},~\bfnm{Louise}\binits{L.}}
(\byear{2004}).
\btitle{An EM algorithm for estimating equations}.
\bjournal{Journal of Computational and Graphical Statistics}
\bvolume{13}
\bpages{48-65}.
\end{barticle}
\endbibitem

\bibitem[\protect\citeauthoryear{Fishkind et~al.}{2013}]{Fishkind-13}
\begin{barticle}[author]
\bauthor{\bsnm{Fishkind},~\bfnm{Donniell~E.}\binits{D.~E.}},
  \bauthor{\bsnm{Sussman},~\bfnm{Daniel~L.}\binits{D.~L.}},
  \bauthor{\bsnm{Vogelstein},~\bfnm{Joshua~T.}\binits{J.~T.}} \AND
  \bauthor{\bsnm{Priebe},~\bfnm{Carey~E.}\binits{C.~E.}}
(\byear{2013}).
\btitle{Consistent adjacency-spectral partitioning for the stochastic block
  model when the model paramters are unknown}.
\bjournal{SIAM Journal on Matrix Analysis and Applications}
\bvolume{34}
\bpages{23-39}.
\end{barticle}
\endbibitem

\bibitem[\protect\citeauthoryear{Fraley and Raftery}{2002}]{Raftery-02}
\begin{barticle}[author]
\bauthor{\bsnm{Fraley},~\bfnm{Chris}\binits{C.}} \AND
  \bauthor{\bsnm{Raftery},~\bfnm{Adrian~E.}\binits{A.~E.}}
(\byear{2002}).
\btitle{Model-based clustering, discriminant analysis, and density estimation}.
\bjournal{Journal of the American Statistical Association}
\bvolume{97}
\bpages{611-631}.
\end{barticle}
\endbibitem

\bibitem[\protect\citeauthoryear{Holland, Askey and Leinhardt}{1983}]{SBM-83}
\begin{barticle}[author]
\bauthor{\bsnm{Holland},~\bfnm{Paul~W.}\binits{P.~W.}},
  \bauthor{\bsnm{Askey},~\bfnm{Kathryn~Blackmond}\binits{K.~B.}} \AND
  \bauthor{\bsnm{Leinhardt},~\bfnm{Samuel}\binits{S.}}
(\byear{1983}).
\btitle{Stochastic blockmodels: First steps}.
\bjournal{Social Networks}
\bvolume{5}
\bpages{109-137}.
\end{barticle}
\endbibitem

\bibitem[\protect\citeauthoryear{Hubert and Arabie}{1985}]{ARI}
\begin{barticle}[author]
\bauthor{\bsnm{Hubert},~\bfnm{Lawrence}\binits{L.}} \AND
  \bauthor{\bsnm{Arabie},~\bfnm{Phipps}\binits{P.}}
(\byear{1985}).
\btitle{Comparing partitions}.
\bjournal{Journal of Classification}
\bvolume{2}
\bpages{193-218}.
\end{barticle}
\endbibitem

\bibitem[\protect\citeauthoryear{Karrer and Newman}{2011}]{Karrer-2011}
\begin{barticle}[author]
\bauthor{\bsnm{Karrer},~\bfnm{Brian}\binits{B.}} \AND
  \bauthor{\bsnm{Newman},~\bfnm{Mark E.~J.}\binits{M.~E.~J.}}
(\byear{2011}).
\btitle{Stochastic blockmodels and community structure in networks}.
\bjournal{Physical Review E}
\bvolume{83}.
\end{barticle}
\endbibitem

\bibitem[\protect\citeauthoryear{Lyzinski et~al.}{2014}]{ASE-14}
\begin{barticle}[author]
\bauthor{\bsnm{Lyzinski},~\bfnm{Vince}\binits{V.}},
  \bauthor{\bsnm{Sussman},~\bfnm{Daniel~L.}\binits{D.~L.}},
  \bauthor{\bsnm{Tang},~\bfnm{Minh}\binits{M.}},
  \bauthor{\bsnm{Athreya},~\bfnm{Avanti}\binits{A.}} \AND
  \bauthor{\bsnm{Priebe},~\bfnm{Carey~E.}\binits{C.~E.}}
(\byear{2014}).
\btitle{Perfect clustering for stochastic blockmodel graphs via adjacency
  spectral embedding}.
\bjournal{Electronic Journal of Statistics}
\bvolume{8}
\bpages{2905-2922}.
\end{barticle}
\endbibitem

\bibitem[\protect\citeauthoryear{Lyzinski et~al.}{2017}]{Lys-2017}
\begin{barticle}[author]
\bauthor{\bsnm{Lyzinski},~\bfnm{Vince}\binits{V.}},
  \bauthor{\bsnm{Tang},~\bfnm{Minh}\binits{M.}},
  \bauthor{\bsnm{Athreya},~\bfnm{Avanti}\binits{A.}},
  \bauthor{\bsnm{Park},~\bfnm{Youngser}\binits{Y.}} \AND
  \bauthor{\bsnm{Priebe},~\bfnm{Carey~E.}\binits{C.~E.}}
(\byear{2017}).
\btitle{Community Detection and Classification in Hierarchical Stochastic
  Blockmodels}.
\bjournal{IEEE Transactions on Network Science and Engineering}
\bvolume{4}
\bpages{13-26}.
\end{barticle}
\endbibitem

\bibitem[\protect\citeauthoryear{Mann and Whitney}{1947}]{Wilcox-rank-sum}
\begin{barticle}[author]
\bauthor{\bsnm{Mann},~\bfnm{Henry~B.}\binits{H.~B.}} \AND
  \bauthor{\bsnm{Whitney},~\bfnm{Donald~R.}\binits{D.~R.}}
(\byear{1947}).
\btitle{On a test of whether one of two random variables is stochastically
  larger than the other.}
\bjournal{Annals of Mathematical Statistics}
\bvolume{18}
\bpages{50-60}.
\end{barticle}
\endbibitem

\bibitem[\protect\citeauthoryear{McLachlan and Krishnan}{2008}]{EM-text}
\begin{bbook}[author]
\bauthor{\bsnm{McLachlan},~\bfnm{Geoffrey~J.}\binits{G.~J.}} \AND
  \bauthor{\bsnm{Krishnan},~\bfnm{Thriyambakam}\binits{T.}}
(\byear{2008}).
\btitle{The EM Algorithm and Extensions},
\bedition{2nd} ed.
\bpublisher{Wiley-Interscience}.
\end{bbook}
\endbibitem

\bibitem[\protect\citeauthoryear{Priebe et~al.}{2017}]{Droso-2017}
\begin{barticle}[author]
\bauthor{\bsnm{Priebe},~\bfnm{Carey~E.}\binits{C.~E.}},
  \bauthor{\bsnm{Park},~\bfnm{Youngser}\binits{Y.}},
  \bauthor{\bsnm{Tang},~\bfnm{Minh}\binits{M.}},
  \bauthor{\bsnm{Athreya},~\bfnm{Avanti}\binits{A.}},
  \bauthor{\bsnm{Lyzinski},~\bfnm{Vince}\binits{V.}} \AND
  \bauthor{\bsnm{Vogelstein},~\bfnm{Joshua~T.}\binits{J.~T.}}
(\byear{2017}).
\btitle{Semiparametric Spectral Modeling of the {\textit{Drosophila}}
  Connectome}.
\bjournal{arXiv}.
\end{barticle}
\endbibitem

\bibitem[\protect\citeauthoryear{Priebe et~al.}{2019}]{Priebe-2019}
\begin{binproceedings}[author]
\bauthor{\bsnm{Priebe},~\bfnm{Carey~E.}\binits{C.~E.}},
  \bauthor{\bsnm{Park},~\bfnm{Youngser}\binits{Y.}},
  \bauthor{\bsnm{Vogelstein},~\bfnm{Joshua~T.}\binits{J.~T.}},
  \bauthor{\bsnm{Conroy},~\bfnm{John~M.}\binits{J.~M.}},
  \bauthor{\bsnm{Lyzinski},~\bfnm{Vince}\binits{V.}},
  \bauthor{\bsnm{Tang},~\bfnm{Minh}\binits{M.}},
  \bauthor{\bsnm{Athreya},~\bfnm{Avanti}\binits{A.}},
  \bauthor{\bsnm{Cape},~\bfnm{Joshua}\binits{J.}} \AND
  \bauthor{\bsnm{Bridgeford},~\bfnm{Eric}\binits{E.}}
(\byear{2019}).
\btitle{On a two-truths phenomenon in spectral graph clustering}.
In \bbooktitle{Proceedings of the National Academy of Sciences of the United
  States of America}.
\bseries{116}
\bvolume{13}
\bpages{5995-6000}.
\end{binproceedings}
\endbibitem

\bibitem[\protect\citeauthoryear{Rubin-Delanchy et~al.}{2020}]{GRDPG}
\begin{barticle}[author]
\bauthor{\bsnm{Rubin-Delanchy},~\bfnm{Patrick}\binits{P.}},
  \bauthor{\bsnm{Cape},~\bfnm{Joshua}\binits{J.}},
  \bauthor{\bsnm{Tang},~\bfnm{Minh}\binits{M.}} \AND
  \bauthor{\bsnm{Priebe},~\bfnm{Carey~E.}\binits{C.~E.}}
(\byear{2020}).
\btitle{A statistical interpretation of spectral embedding: the generalised
  random dot product graph}.
\bjournal{arXiv}.
\end{barticle}
\endbibitem

\bibitem[\protect\citeauthoryear{Schwarz}{1978}]{BIC}
\begin{barticle}[author]
\bauthor{\bsnm{Schwarz},~\bfnm{Gideon~E.}\binits{G.~E.}}
(\byear{1978}).
\btitle{Estimating the dimension of a model}.
\bjournal{The Annals of Statistics}
\bvolume{6}
\bpages{464-464}.
\end{barticle}
\endbibitem

\bibitem[\protect\citeauthoryear{Scrucca et~al.}{2016}]{Raftery-16}
\begin{barticle}[author]
\bauthor{\bsnm{Scrucca},~\bfnm{Luca}\binits{L.}},
  \bauthor{\bsnm{Fop},~\bfnm{Michael}\binits{M.}},
  \bauthor{\bsnm{Murphy},~\bfnm{T.~Brendan}\binits{T.~B.}} \AND
  \bauthor{\bsnm{Raftery},~\bfnm{Adrian~E.}\binits{A.~E.}}
(\byear{2016}).
\btitle{mclust 5: Clustering, classification, and density estimation using
  {G}aussian finite mixture models}.
\bjournal{The R Journal}
\bvolume{8}
\bpages{289-317}.
\end{barticle}
\endbibitem

\bibitem[\protect\citeauthoryear{Sussman et~al.}{2012}]{Sussman-2012}
\begin{barticle}[author]
\bauthor{\bsnm{Sussman},~\bfnm{Daniel~L.}\binits{D.~L.}},
  \bauthor{\bsnm{Tang},~\bfnm{Minh}\binits{M.}},
  \bauthor{\bsnm{Fishkind},~\bfnm{Donniell~E.}\binits{D.~E.}} \AND
  \bauthor{\bsnm{Priebe},~\bfnm{Carey~E.}\binits{C.~E.}}
(\byear{2012}).
\btitle{A consistent adjacency spectral embedding for stochastic blockmodel
  graphs}.
\bjournal{Journal of the American Statistical Association}
\bvolume{107}
\bpages{1119-1128}.
\end{barticle}
\endbibitem

\bibitem[\protect\citeauthoryear{Tang and Priebe}{2018}]{Tang-18}
\begin{barticle}[author]
\bauthor{\bsnm{Tang},~\bfnm{Minh}\binits{M.}} \AND
  \bauthor{\bsnm{Priebe},~\bfnm{Carey~E.}\binits{C.~E.}}
(\byear{2018}).
\btitle{Limit theorems for eigenvectors of the normalized Laplacian for random
  graphs}.
\bjournal{The Annals of Statistics}
\bvolume{46}
\bpages{2360-2415}.
\end{barticle}
\endbibitem

\bibitem[\protect\citeauthoryear{Xie}{2021}]{xie-21}
\begin{barticle}[author]
\bauthor{\bsnm{Xie},~\bfnm{Fangzheng}\binits{F.}}
(\byear{2021}).
\btitle{Entrywise limit theorems of eigenvectors and their one-step refinement
  for sparse random graphs}.
\bjournal{arXiv}.
\end{barticle}
\endbibitem

\bibitem[\protect\citeauthoryear{Xie and Xu}{2019}]{xie-19}
\begin{barticle}[author]
\bauthor{\bsnm{Xie},~\bfnm{Fangzheng}\binits{F.}} \AND
  \bauthor{\bsnm{Xu},~\bfnm{Yanxun}\binits{Y.}}
(\byear{2019}).
\btitle{Efficient estimation for Random Dot Product Graphs via a one-step
  procedure}.
\bjournal{arXiv}.
\end{barticle}
\endbibitem

\bibitem[\protect\citeauthoryear{Yang et~al.}{2021}]{yang-21}
\begin{barticle}[author]
\bauthor{\bsnm{Yang},~\bfnm{Congyuan}\binits{C.}},
  \bauthor{\bsnm{Priebe},~\bfnm{Carey~E.}\binits{C.~E.}},
  \bauthor{\bsnm{Park},~\bfnm{Youngser}\binits{Y.}} \AND
  \bauthor{\bsnm{Marchette},~\bfnm{David~J.}\binits{D.~J.}}
(\byear{2021}).
\btitle{Simultaneous dimensionality and complexity model selection for spectral
  graph clustering}.
\bjournal{Journal of Computational and Graphical Statistics}
\bvolume{30}
\bpages{422--441}.
\end{barticle}
\endbibitem

\bibitem[\protect\citeauthoryear{Young and Scheinerman}{2007}]{Young-07}
\begin{binproceedings}[author]
\bauthor{\bsnm{Young},~\bfnm{Stephan~J.}\binits{S.~J.}} \AND
  \bauthor{\bsnm{Scheinerman},~\bfnm{Edward}\binits{E.}}
(\byear{2007}).
\btitle{Random dot product graph models for social networks}.
In \bbooktitle{Proceedings of the 5th international conference on algorithms
  and models for the web-graph}
\bpages{138-149}.
\end{binproceedings}
\endbibitem

\end{thebibliography}

\appendix
\renewcommand{\thesection}{\Roman{section}}

\section{The Expectation-Solution Algorithm for the General Incomplete Data Setting}\label{app:appendix1}
We model our treatment of the ES algorithm on that found in \cite{EM-text}.

Suppose $\textbf X \sim f_{\boldsymbol\Psi}$ is observed with realization \textbf{x}, $\boldsymbol\Psi \in \mathbbm R^p$, and there exists a natural unobserved extension of the observed data $\textbf Y = (\textbf X, \textbf Z)$. Let
\begin{equation*}
U_c(\textbf Y, \boldsymbol\Psi) = \boldsymbol 0
\end{equation*}
be the $p$-dimensional complete-data estimating equation that could be solved if $\textbf Y$ were totally observable. To apply ES, we first re-express the complete-data estimating equation in terms of a linear function of a $q$-dimensional vector function $S(\textbf Y)$, a $(p\times q)$-dimensional matrix function $\boldsymbol C_{\boldsymbol\Psi}$, and a $p$-dimensional vector function $b_{\boldsymbol\Psi}(\textbf X)$:
\begin{align*}
U_c(\textbf Y, \boldsymbol\Psi) &= U^{(1)}(\textbf X, S(\textbf Y), \boldsymbol\Psi) \\
&= \sum_{j = 1}^q c_j(\boldsymbol\Psi)S_j(\textbf Y) + b_{\boldsymbol\Psi}(\textbf X) \\
&= \boldsymbol{C_\Psi} S(\textbf Y) + b_{\boldsymbol\Psi}(\textbf X) \\
&= \boldsymbol 0,
\end{align*}
where $c_j(\boldsymbol\Psi)$ is the $j$th column of $\boldsymbol C_{\boldsymbol\Psi}$. Here $S(\textbf Y)$ is the complete data summary statistic. 

The algorithm is:

\noindent\fbox{\parbox{\textwidth}{
\noindent\fbox{\textbf{The ES Algorithm for the General Incomplete Data Setting}}

1. Initialize $\boldsymbol{\Psi^*} = \boldsymbol{\Psi_0}$.

2. \textbf{E-Step:} Compute $h(\boldsymbol\Psi | \boldsymbol{\Psi^*}) := \mathbbm E_{\boldsymbol{\Psi^*}} \lbrack S(\textbf Y) | \textbf{X} = \textbf{x} \rbrack$.

3. \textbf{S-Step:} Find $\boldsymbol{\hat\Psi}$ that solves $\textbf{A}_{\boldsymbol{\hat\Psi}}h(\boldsymbol{\hat\Psi} |\boldsymbol{\Psi^*}) + b_{\boldsymbol{\hat\Psi}}(\textbf{x}) = \boldsymbol 0$.

4. Take $\boldsymbol{\Psi^*} = \boldsymbol{\hat\Psi}$.

5. Repeat steps 2--4 until some convergence criterion is satisfied.
}
}

\section{Comparison of EM and ES to $K$-Means}\label{app:appendix2}

In our simulations we also compared the clustering performance of EM and ES against that of $K$-means in models 1--4 and the brain connectome. We present the results in terms of 95\% confidence intervals for the median difference in ARI, but due to the presence of multiple ties in the homogeneous balanced affinity settings we elected to use the sign test with a two-sided alternative. For $\ell = A, L$ as in section 4 and $Z = M, S$ we define
\begin{equation*}
	\Delta_\ell^{KZ} := ARI_{KM\circ\ell SE}-ARI_{EZ\circ\ell SE}.
\end{equation*}
We embolden all entries indicating strict, significant improvement of our ES algorithms over $K$-means.



\begin{table}[h!]
\caption{Model 1 Mixture}
\tiny
\centering
\begin{tabular}{l|llll}
$n$ & 95\% CI for med$(\Delta_A^{KM})$ & 95\% CI for med$(\Delta_L^{KM})$ & 95\% CI for med$(\Delta_A^{KS})$ & 95\% CI for med$(\Delta_L^{KS})$ \\ \hline
200 & (0.0167, 0.0493)                 & (0,0.0220)                       & (0,0.0048)                       & (0, 0.015)                       \\
300 & (0, 0.0122)                      & (0, 0.0119)                      & (0, 0)                           & (0, 0)                           \\
400 & (0, 0)                           & (0, 0)                           & (0, 0)                           & (0, 0)                           \\
500 & (0, 0)                           & (0, 0)                           & (0, 0)                           & (0, 0)                           \\
600 & (0, 0)                           & (0, 0)                           & (0, 0)                           & (0, 0)                           \\
700 & (0, 0)                           & (0, 0)                           & (0, 0)                           & (0, 0)                           \\
800 & (0, 0)                           & (0, 0)                           & (0, 0)                           & (0, 0)                           \\
900 & (0, 0)                           & (0, 0)                           & (0, 0)                           & (0, 0)                          
\end{tabular}
\end{table}


\begin{table}[h!]
\caption{Model 2 Mixture}
\tiny
\centering
\begin{tabular}{l|llll}
$n$ & 95\% CI for med$(\Delta_A^{KM})$ & 95\% CI for med$(\Delta_L^{KM})$ & 95\% CI for med$(\Delta_A^{KS})$ & 95\% CI for med$(\Delta_L^{KS})$ \\ \hline
200 & (0.0407, 0.0915)                 & (0.0139, 0.0439)                 & (0.0127, 0.0761)                 & (0.0140, 0.0774)                 \\
300 & (0.0098, 0.0313)                 & (0.0061, 0.0197)                 & (1e-7, -0.0190)                  & (0, 0.0105)                      \\
400 & (0.0131, 0.0168)                 & (-2e-7, 0.0106)                  & (-1e-7, 0.0023)                  & (0, 0.0023)                      \\
500 & (-0.0067, 0.0134)                & (-0.0066, 0.0068)                & (0, 2e-7)                        & (0, 0.0066)                      \\
600 & (-2e-8, 0.0059)                  & (0, 0.0059)                      & (0, 5e-9)                        & (0, 0.0018)                      \\
700 & (0, 0.0053)                      & (0, 0.0051)                      & (0, 0)                           & (0, 0)                           \\
800 & (0, 0.0046)                      & (0, 0.0046)                      & (0, 4e-8)                        & (0, 0)                           \\
900 & (-2e-9, 0.0041)                  & (-7e-10, 0.0013)                 & (0, 0)                           & (0, 0)                          
\end{tabular}
\end{table}


\begin{table}[h!]
\caption{Model 3 Mixture}
\tiny
\centering
\begin{tabular}{l|llll}
$n$ & 95\% CI for med$(\Delta_A^{KM})$ & 95\% CI for med$(\Delta_L^{KM})$ & 95\% CI for med$(\Delta_A^{KS})$ & 95\% CI for med$(\Delta_L^{KS})$ \\ \hline
200 & (-0.1070, -0.0537)               & (-0.1450, -0.0964)               &  \textbf{(-0.0256, -8e-5)}                 &  \textbf{(-0.026, -0.0080)}                \\
300 & (-0.1756, -0.1317)               & (-0.2068, -0.1662)               &  \textbf{(-0.0998, -0.0318)}               &  \textbf{(-0.1273, -0.0888)}               \\
400 & (-0.2019, -0.1712)               & (-0.2215, -0.1966)               &  \textbf{(-0.1682, -0.1204)}               &  \textbf{(-0.1762, -0.1242)}               \\
500 & (-0.2306, -0.2002)               & (-0.2476, -0.2263)               &  \textbf{(-0.2288, -0.1986)}               &  \textbf{(-0.2370, -0.2103)}               \\
600 & (-0.2514, -0.2305)               & (-0.2671, -0.2437)               &  \textbf{(-0.2547, -0.2296)}               &  \textbf{(-0.2656, -0.2434)}               \\
700 & (-0.2573, -0.2380)               & (-0.2772, -0.2555)               &  \textbf{(-0.2657, -0.2386)}               &  \textbf{(-0.2755, -0.2577)}               \\
800 & (-0.2534, -0.2388)               & (-0.2716, -0.2562)               &  \textbf{(-0.2619, -0.2370)}               &  \textbf{(-0.2760, -0.2574)}               \\
900 & (-0.2495, -0.2370)               & (-0.2686, -0.2586)               &  \textbf{(-0.2528, -0.2426)}               &  \textbf{(-0.2722, 0.2577)}               
\end{tabular}
\end{table}


\begin{table}[h!]
\caption{Model 4 Mixture}
\tiny
\centering
\begin{tabular}{l|llll}
$n$ & 95\% CI for med$(\Delta_A^{KM})$ & 95\% CI for med$(\Delta_L^{KM})$ & 95\% CI for med$(\Delta_A^{KS})$ & 95\% CI for med$(\Delta_L^{KS})$ \\ \hline
200 & (-0.0543, 0.0101)                & (-0.0777, -0.0244)               &  \textbf{(-0.0066, 0.0428)}                &  \textbf{(-0.0099, -0.0077)}               \\
300 & (-0.1207, -0.0883)               & (-0.1306, -0.0908)               &  \textbf{(-0.0316, 0.1010)}                &  \textbf{(-0.0606, -0.0346)}               \\
400 & (-0.1704, -0.1367)               & (-0.1772, -0.1520)               &  \textbf{(-0.0887, -0.0466)}               &  \textbf{(-0.1018, -0.0688)}               \\
500 & (-0.2116, -0.1890)               & (-0.2191, -0.1912)               &  \textbf{(-0.1310, -0.943)}                &  \textbf{(-0.1502, -0.0961)}               \\
600 & (-0.2270, -0.1970)               & (-0.2412, -0.2150)               &  \textbf{(-0.1992, -0.1556)}               &  \textbf{(-0.2252, -0.1858)}               \\
700 & (-0.2436, -0.2149)               & (-0.2607, -0.2342)               &  \textbf{(-0.2345, -0.2051)}               &  \textbf{(-0.2505, -0.2301)}               \\
800 & (-0.2476, -0.2252)               & (-0.2678, -0.2462)               &  \textbf{(-0.2420, -0.2214)}               &  \textbf{(-0.2625, -0.2399)}               \\
900 & (-0.2512, -0.2372)               & (-0.2777, -0.2572)               &  \textbf{(-0.2540, -0.2363)}               &  \textbf{(-0.2785, -0.2611)}              
\end{tabular}
\end{table}


\begin{table}[h!]
\caption{Model 1 SBM}
\tiny
\centering
\begin{tabular}{l|llll}
$n$ & 95\% CI for med$(\Delta_A^{KM})$ & 95\% CI for med$(\Delta_L^{KM})$ & 95\% CI for med$(\Delta_A^{KS})$ & 95\% CI for med$(\Delta_L^{KS})$ \\ \hline
200 & (0.0148, 0.0530)                 & (7e-6, 0.0257)                   & (1e-5, 0.0148)                   & (0, 0.0026)                      \\
300 & (0, 0.0227)                      & (0, 0.0113)                      & (0, 0)                           & (0, 0)                           \\
400 & (0, 0.0088)                      & (0, 0.0086)                      & (0, 0)                           & (0, 0)                           \\
500 & (0, 0)                           & (0, 3e-9)                        & (0, 0)                           & (0, 0)                           \\
600 & (0, 0)                           & (0, 0.0020)                      & (0, 0)                           & (0, 0)                           \\
700 & (0, 0)                           & (0, 0)                           & (0, 0)                           & (0, 0)                           \\
800 & (0, 0)                           & (0, 0)                           & (0, 0)                           & (0, 0)                           \\
900 & (0, 0)                           & (0, 0)                           & (0, 0)                           & (0, 0)                          
\end{tabular}
\end{table}


\begin{table}[h!]
\caption{Model 2 SBM}
\tiny
\centering
\begin{tabular}{l|llll}
$n$ & 95\% CI for med$(\Delta_A^{KM})$ & 95\% CI for med$(\Delta_L^{KM})$ & 95\% CI for med$(\Delta_A^{KS})$ & 95\% CI for med$(\Delta_L^{KS})$ \\ \hline
200 & (0.0033, 0.0135)                 & (0.0010, 0.0109)                 & (0.0094, 0.0265)                 & (0, 0.0015)                      \\
300 & (0.0235, 0.0510)                 & (0.0073, 0.0216)                 & (0.0131, 0.0830)                 & (0, 0.0008)                      \\
400 & (0.0137, 0.0430)                 & (5e-6, 0.0186)                   & (-8e-7, 0.0065)                  & (0, 5e-6)                        \\
500 & (0.0122, 0.0325)                 & (0.0038, 0.0115)                 & (0, 0.0054)                      & (-0.0014, 9e-7)                  \\
600 & (-6e-6, 0.0098)                  & (1e-7, 0.0055)                   & (-1e-6, 8e-8)                    & (-0.0051, 7e-8)                  \\
700 & (0, 0.0063)                      & (0, 0.0047)                      & (-4e-8, 2e-7)                    & (-0.0014, 2e-7)                  \\
800 & (0, 0.0087)                      & (-1e-7, 0.0044)                  & (0, 0)                           & (-2e-9, 0)                       \\
900 & (0, 0.0079)                      & (0, 0.0040)                      & (0,0)                            & (0, 5e-9)                       
\end{tabular}
\end{table}


\begin{table}[h!]
\caption{Model 3 SBM}
\tiny
\centering
\begin{tabular}{l|llll}
$n$ & 95\% CI for med$(\Delta_A^{KM})$ & 95\% CI for med$(\Delta_L^{KM})$ & 95\% CI for med$(\Delta_A^{KS})$ & 95\% CI for med$(\Delta_L^{KS})$ \\ \hline
200 & (-0.1061, -0.0559)               & (-0.1834, -0.1487)               &  \textbf{(-0.1552, -0.0954)}               &  \textbf{(-0.1868, -0.1220)}               \\
300 & (-0.2065, -0.1188)               & (-0.2767, -0.2429)               &  \textbf{(-0.2323, -0.1825)}               &  \textbf{(-0.2796, -0.2473)}               \\
400 & (-0.3298, -0.2676)               & (-0.3666, -0.3380)               &  \textbf{(-0.3401, -0.2991)}               &  \textbf{(-0.3709, -0.3437)}               \\
500 & (-0.3872, -0.3664)               & (-0.4409, -0.4039)               &  \textbf{(-0.3981, -0.3631)}               &  \textbf{(-0.4373, -0.4038)}               \\
600 & (-0.4284, -0.4022)               & (-0.4673, -0.4512)               &  \textbf{(-0.4409, -0.4151)}               &  \textbf{(-0.4816, -0.4551)}               \\
700 & (-0.4900, -0.4617)               & (-0.5192, -0.4958)               &  \textbf{(-0.4835, -0.4501)}               &  \textbf{(-0.5214, -0.4889)}               \\
800 & (-0.5236, -0.4878)               & (-0.5515, -0.5234)               &  \textbf{(-0.5133, -0.4663)}               &  \textbf{(-0.5452, -0.5111)}               \\
900 & (-0.5145, -0.4678)               & (-0.5365, -0.5034)               &  \textbf{(-0.4927, -0.4568)}               &  \textbf{(-0.5307, -0.4972)}              
\end{tabular}
\end{table}


\begin{table}[h!]
\caption{Model 4 SBM}
\tiny
\centering
\begin{tabular}{l|llll}
$n$ & 95\% CI for med$(\Delta_A^{KM})$ & 95\% CI for med$(\Delta_L^{KM})$ & 95\% CI for med$(\Delta_A^{KS})$ & 95\% CI for med$(\Delta_L^{KS})$ \\ \hline
200 & (-0.0781, -0.0312)               & (-0.1298, -0.0814)               &  \textbf{(-0.0050, -5e-5)}                 &  \textbf{(-0.1049, -0.0718)}               \\
300 & (-0.1356, -0.0636)               & (-0.2158, -0.1760)               &  \textbf{(-0.0233, -0.0090)}               &  \textbf{(-0.2061, -0.1693)}               \\
400 & (-0.1975, -0.1219)               & (-0.2891, -0.2402)               &  \textbf{(-0.0916, -0.0621)}               &  \textbf{(-0.2826, -0.2527)}               \\
500 & (-0.2809, -0.2366)               & (-0.3505, -0.3271)               & \textbf{(-0.2265, -0.1744)}               &  \textbf{(-0.3466, -0.3235)}               \\
600 & (-0.3477, -0.3134)               & (-0.4117, -0.3841)               &  \textbf{(-0.3084, -0.2622)}               &  \textbf{(-0.4078, -0.3756)}               \\
700 & (-0.4164, -0.3799)               & (-0.4611, -0.4339)               &  \textbf{(-0.3875, -0.3591)}               &  \textbf{(-0.4630, -0.4344)}               \\
800 & (-0.4417, -0.4144)               & (-0.4962, -0.4755)               &  \textbf{(-0.4413, -0.4067)}               &  \textbf{(-0.5070, -0.4767)}               \\
900 & (-0.4970, -0.4754)               & (-0.5504, -0.5230)               &  \textbf{(-0.4965, -0.4661)}               &  \textbf{(-0.5469, -0.5286)}              
\end{tabular}
\end{table}


\begin{table}[h!]
\caption{Brain Connectome SBM}
\tiny
\centering
\begin{tabular}{l|llll}
$n$  & 95\% CI for med$(\Delta_A^{KM})$ & 95\% CI for med$(\Delta_L^{KM})$ & 95\% CI for med$(\Delta_A^{KS})$ & 95\% CI for med$(\Delta_L^{KS})$ \\ \hline
500  & (-0.5238, -0.5044)               & (-0.5679, -0.5493)               &  \textbf{(-0.5135, -0.4820)}               &  \textbf{(-0.5643, -0.5336)}               \\
600  & (-0.5514, -0.5200)               & (-0.5909, -0.5672)               &  \textbf{(-0.5322, -0.5055)}               &  \textbf{(-0.5811, -0.5472)}               \\
700  & (-0.5676, -0.5432)               & (-0.6148, -0.5929)               &  \textbf{(-0.5453, -0.5199)}               &  \textbf{(-0.5953, -0.5665)}               \\
800  & (-0.5430, -0.5032)               & (-0.5907, -0.5580)               &  \textbf{(-0.5229, -0.4929)}               &  \textbf{(-0.5722, -0.5447)}               \\
900  & (-0.5442, -0.4975)               & (-0.5824, -0.5468)               &  \textbf{(-0.5303, -0.4956)}               &  \textbf{(-0.5744, -0.5439)}               \\
1000 & (-0.5119, -0.4769)               & (-0.5498, -0.5247)               &  \textbf{(-0.5029, -0.4747)}               &  \textbf{(-0.5461, -0.5212)}               \\
1100 & (-0.4952, -0.4709)               & (-0.5363, -0.5170)               &  \textbf{(-0.4938, -0.4682)}               &  \textbf{(-0.5362, -0.5093)}               \\
1200 & (-0.4701, -0.4462)               & (-0.5155, -0.4904)               &  \textbf{(-0.4719, -0.4489)}               &  \textbf{(-0.5123, -0.4914)}              
\end{tabular}
\end{table}
\FloatBarrier

\section{Additional Rank-Deficient Simulations}\label{app:rank_deficient_appendix}

\begin{figure}[h!]
	\centering
	\includegraphics[width = \textwidth]{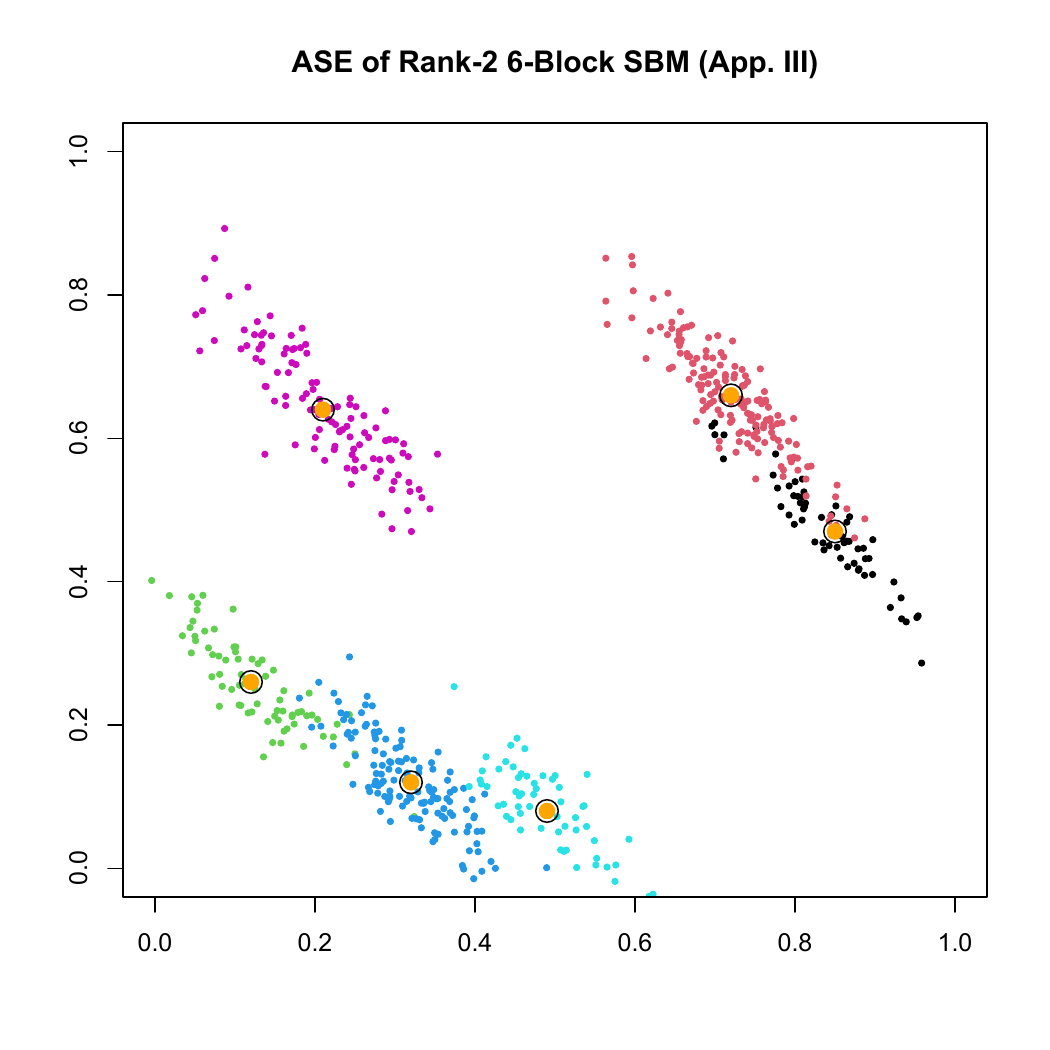}
	\caption{\label{fig:low_rank_ase_2}Example two-dimensional ASE of a random graph generated from the first rank-deficient SBM in Appendix \ref{app:rank_deficient_appendix}. The smaller points have been colored according to block membership of the corresponding nodes in the original graph. The larger orange points encircled with a black outline denote the true latent positions, i.e., the rows of $\textbf{X}$.}
\end{figure}

\begin{figure}[h!]
	\centering
	\includegraphics[width = \textwidth]{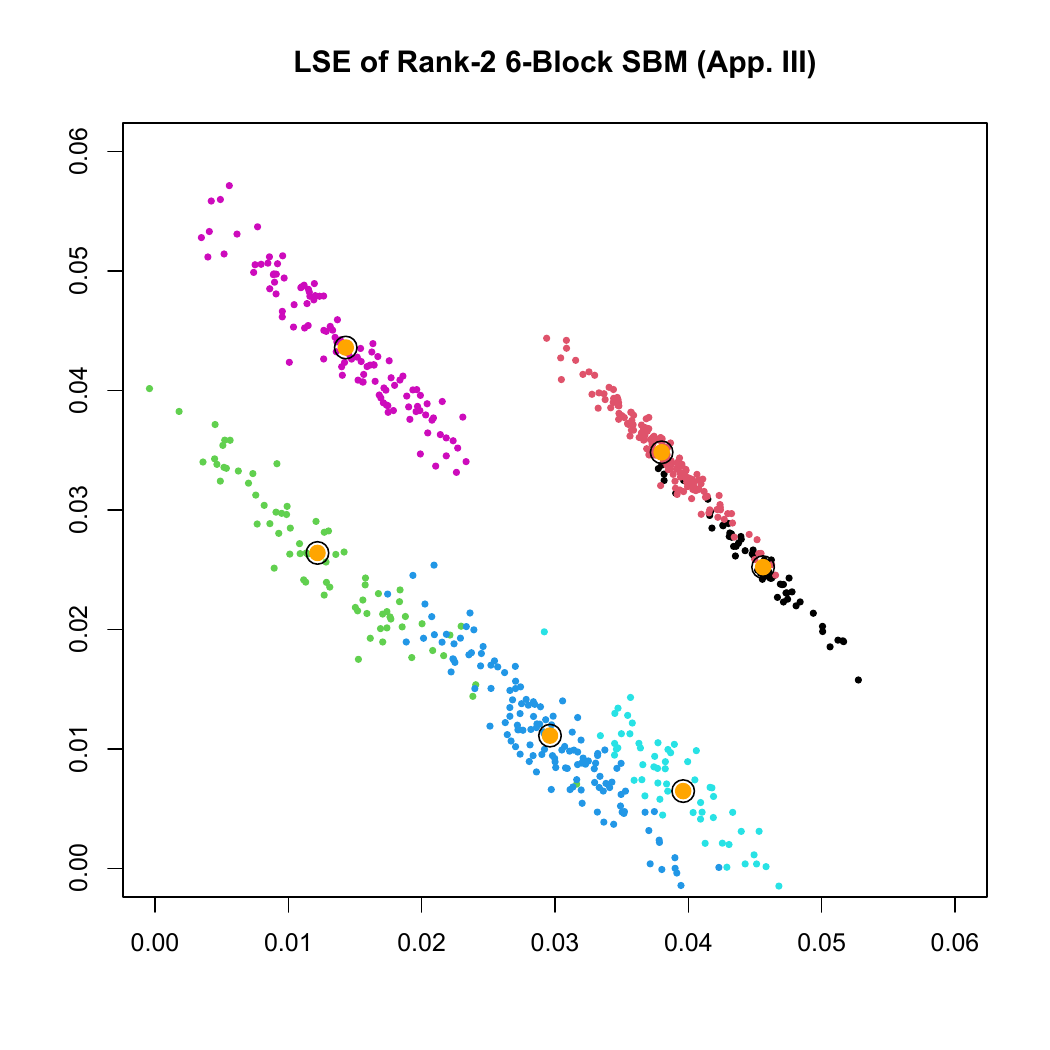}
	\caption{\label{fig:low_rank_lse_2}Example two-dimensional LSE of a random graph generated from the first rank-deficient SBM in Appendix \ref{app:rank_deficient_appendix}. The smaller points have been colored according to block membership of the corresponding nodes in the original graph. The larger orange points encircled with a black outline denote the scaled latent positions, i.e., the centers of the limiting C-GMM.}
\end{figure}

\begin{figure}[h!]
	\centering
	\includegraphics[width = \textwidth]{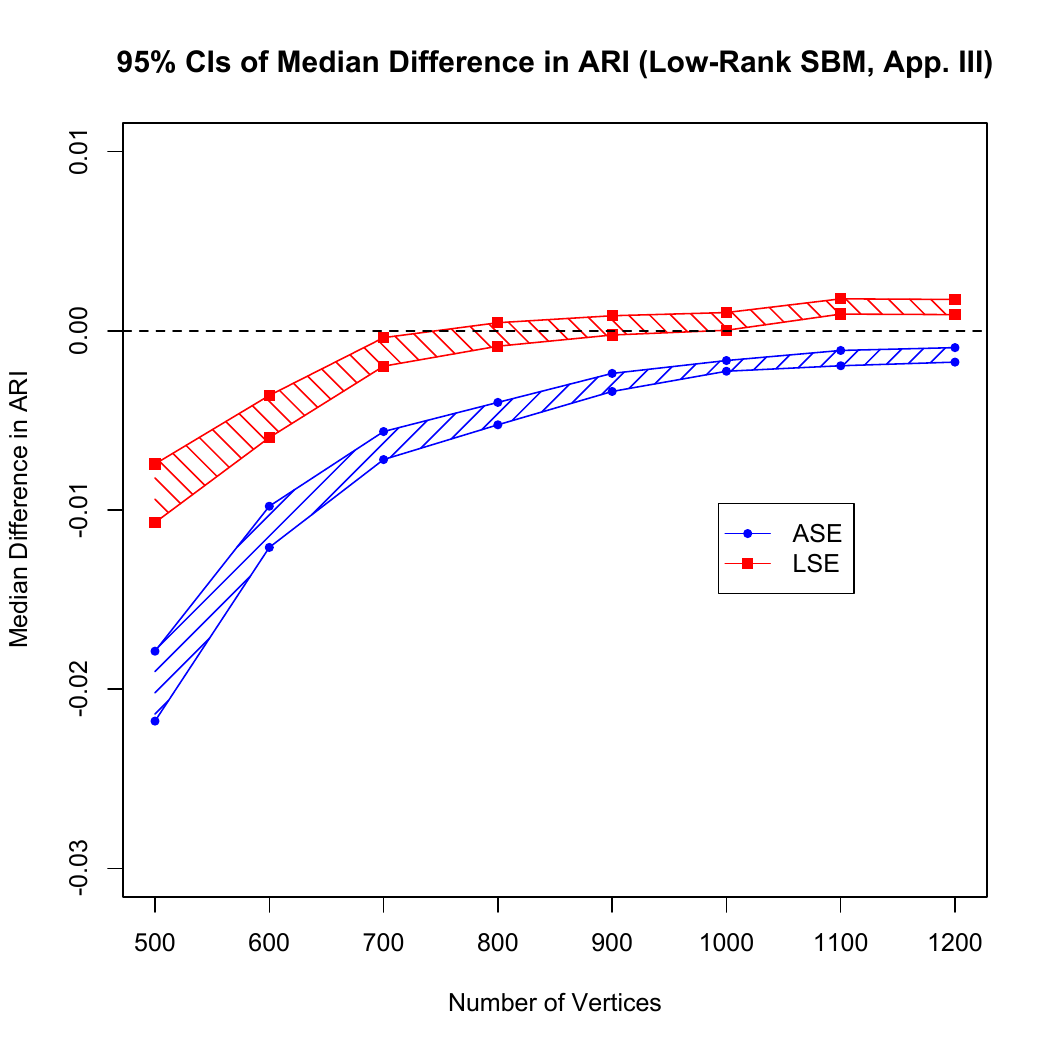}
	\caption{\label{fig:low_rank_results_2}Simulation results for the first rank-deficient SBM in Appendix \ref{app:rank_deficient_appendix}.}
\end{figure}

We include the results of additional simulations in two randomly generated rank-deficient SBM settings analogous to those presented in Subsection \ref{low_rank_sims}. First consider the rank-2 g-block SBM with proportion vector
\begin{equation*}
	\boldsymbol\pi = (.10,.27,.12,.22,.10,.19)
\end{equation*}
and canonical latent position matrix
\begin{equation*}
	\textbf{x} = \begin{bmatrix}
		0.85& 0.47\\
		0.72& 0.66\\
		0.12& 0.26\\
		0.32& 0.12\\
		0.49& 0.08\\
		0.21& 0.64
	\end{bmatrix}
\end{equation*}
which corresponds to the block probability matrix
\begin{equation*}
	\textbf{B} = \begin{bmatrix}
		0.9434 &0.9222 &0.2242 &0.3284 &0.4541 &0.4793\\
		0.9222 &0.9540 &0.2580 &0.3096 &0.4056 &0.5736\\
		0.2242 &0.2580 &0.0820 &0.0696 &0.0796 &0.1916\\
		0.3284 &0.3096 &0.0696 &0.1168 &0.1664 &0.1440\\
		0.4541 &0.4056 &0.0796 &0.1664 &0.2465 &0.1541\\
		0.4793 &0.5736 &0.1916 &0.1440 &0.1541 &0.4537
	\end{bmatrix}.
\end{equation*}
The ASE and LSE of one particular instance ($n=600$) are presented in Figures \ref{fig:low_rank_ase_2}--\ref{fig:low_rank_lse_2}. We present the results in Figure \ref{fig:low_rank_results_2}; $ES\circ ASE$ maintains its dominance over $EM\circ ASE$ here, but $ES\circ LSE$ appears to falter against $EM\circ LSE$ as $n$ increases.

\FloatBarrier

\begin{figure}[h!]
	\centering
	\includegraphics[width = \textwidth]{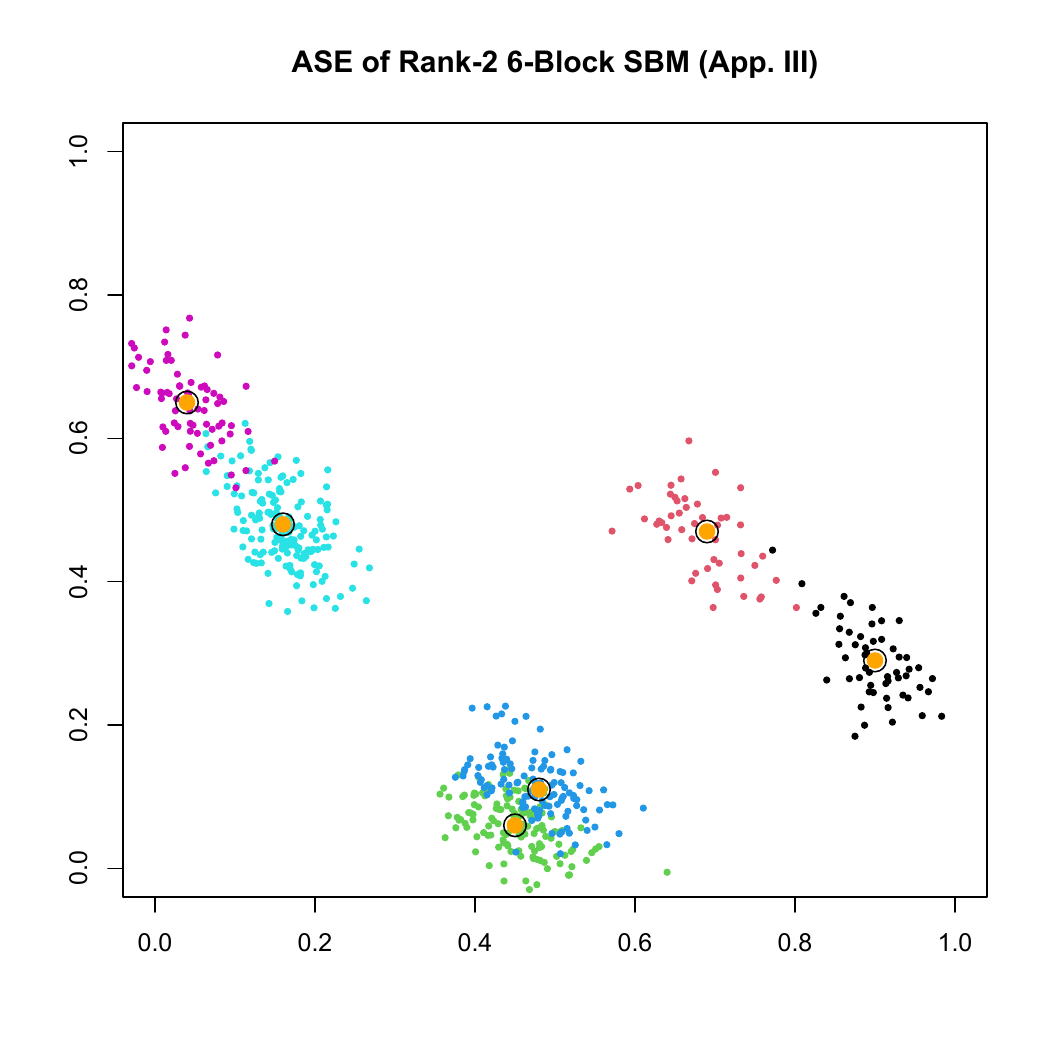}
	\caption{\label{fig:low_rank_ase_3}Example two-dimensional ASE of a random graph generated from the second rank-deficient SBM in Appendix \ref{app:rank_deficient_appendix}. The smaller points have been colored according to block membership of the corresponding nodes in the original graph. The larger orange points encircled with a black outline denote the true latent positions, i.e., the rows of $\textbf{X}$.}
\end{figure}

\begin{figure}[h!]
	\centering
	\includegraphics[width = \textwidth]{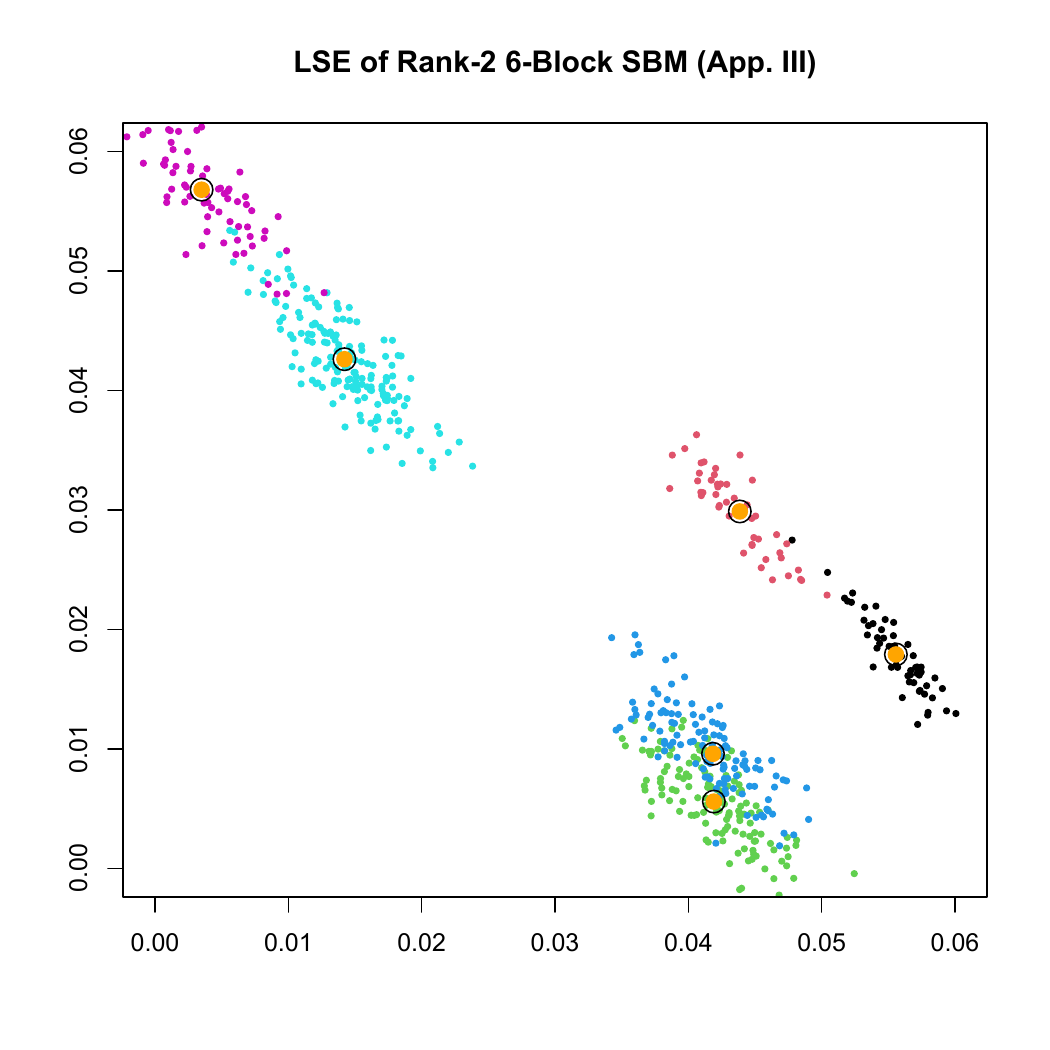}
	\caption{\label{fig:low_rank_lse_3}Example two-dimensional LSE of a random graph generated from the second rank-deficient SBM in Appendix \ref{app:rank_deficient_appendix}. The smaller points have been colored according to block membership of the corresponding nodes in the original graph. The larger orange points encircled with a black outline denote the scaled latent positions, i.e., the centers of the limiting C-GMM.}
\end{figure}

\begin{figure}[h!]
	\centering
	\includegraphics[width = \textwidth]{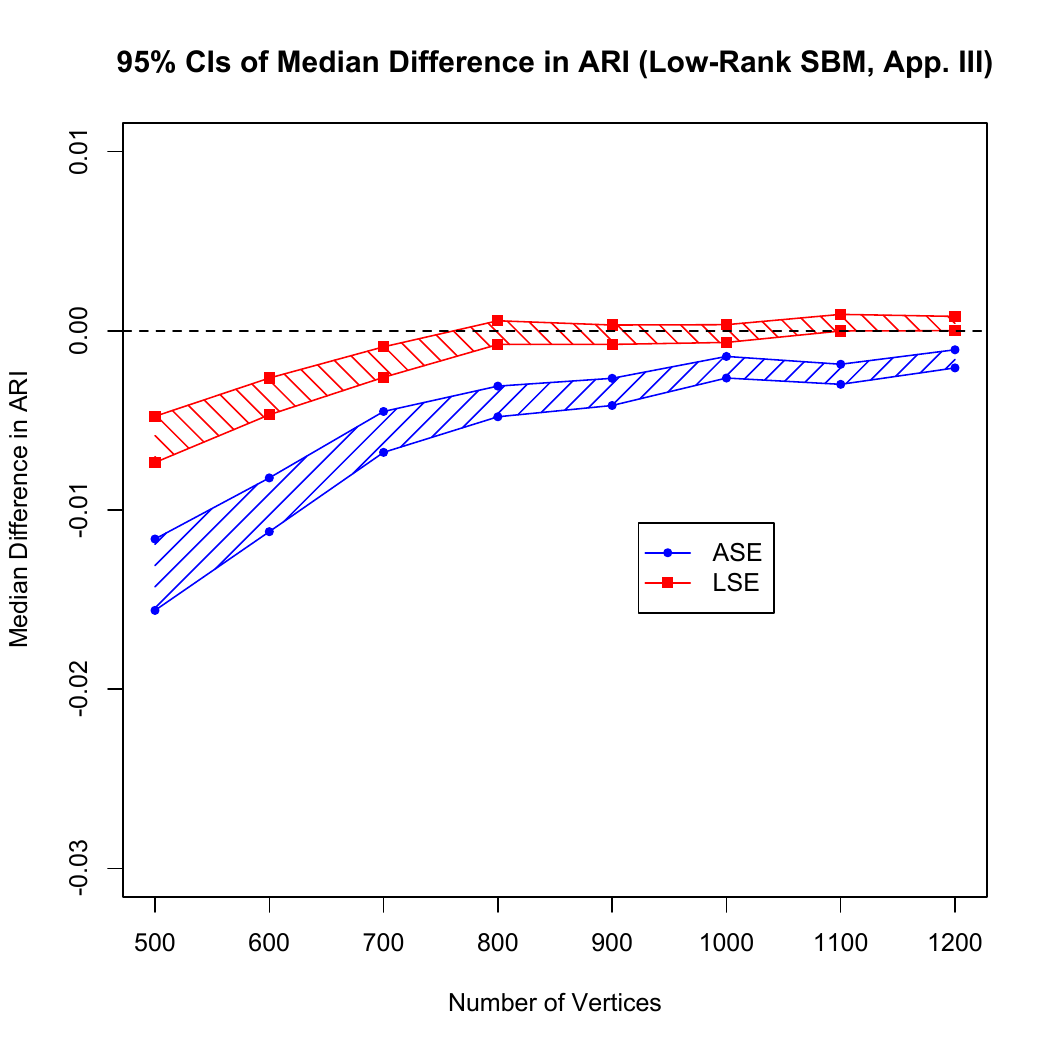}
	\caption{\label{fig:low_rank_results_3}Simulation results for the second rank-deficient SBM in Appendix \ref{app:rank_deficient_appendix}.}
\end{figure}

Now consider another rank-2 g-block SBM with proportion vector
\begin{equation*}
	\boldsymbol\pi = (0.09, 0.08, 0.23, 0.20, 0.28, 0.12)
\end{equation*}
and canonical latent position matrix
\begin{equation*}
	\textbf{x} = \begin{bmatrix}
		0.90 &0.29\\
		0.69 &0.47\\
		0.45 &0.06\\
		0.48 &0.11\\
		0.16 &0.48\\
		0.04 &0.65
	\end{bmatrix}
\end{equation*}
which corresponds to the block probability matrix
\begin{equation*}
	\textbf{B} = \begin{bmatrix}
		0.8941 &0.7573 &0.4224 &0.4639 &0.2832 &0.2245\\
		0.7573 &0.6970 &0.3387 &0.3829 &0.3360 &0.3331\\
		0.4224 &0.3387 &0.2061 &0.2226 &0.1008 &0.0570\\
		0.4639 &0.3829 &0.2226 &0.2425 &0.1296 &0.0907\\
		0.2832 &0.3360 &0.1008 &0.1296 &0.2560 &0.3184\\
		0.2245 &0.3331 &0.0570 &0.0907 &0.3184 &0.4241
	\end{bmatrix}.
\end{equation*}
The ASE and LSE of one particular instance ($n=600$) are presented in Figures \ref{fig:low_rank_ase_3}--\ref{fig:low_rank_lse_3}. We present the results in Figure \ref{fig:low_rank_results_3}, which are similar to those obtained in the previous setting.

\end{document}